%% file: text.tex
\def\mib#1{\mbox{\boldmath $#1$}} 
\documentclass{appolb} 
\usepackage{epsfig} 

\def\mib#1{\mbox{\boldmath $#1$}} 
\begin{document} 
\title{Radiative Corrections to Neutrino-Nucleon  Quasielastic Scattering %
\\
\vskip0.2cm
{\large {\it Dedicated to the Memory of  Jan Kwieci{\' n}ski}} 
}
\author{Masataka Fukugita 
\\ 
\address{Institute for Cosmic Ray Research, University of Tokyo 
\\ 
Kashiwa  277-8582, Japan} 
\and 
Takahiro Kubota 
\\ 
\address{Graduate School of Science, Osaka University 
\\ 
Toyonaka, Osaka 560-0043, Japan} 
} 
\maketitle 
\begin{abstract} 
Full one-loop radiative corrections are calculated for 
low energy neutrino-nucleon quasi-elastic scattering, 
$\bar \nu _{e}+p \longrightarrow e^{+} + n$, 
which involves both Fermi and Gamow-Teller 
transitions, in the static limit of nucleons. 
The calculation is performed for both 
angular independent and dependent parts. 
We separate the corrections into the outer 
and inner parts {\` a} la Sirlin. The outer part 
is infrared and ultraviolet finite, and 
involves the positron velocity. 
The calculation of the outer part is straightforward, but 
that of the inner part requires a scrutiny concerning the 
continuation of the long-distance hadronic calculation 
to the short-distance quark treatment and the 
dependence on the model of hadron structure. We show 
that the logarithmically divergent parts do not depend on the 
structure of hadrons not only for the Fermi part, 
but also for the Gamow-Teller part. This observation enables us 
to deal with the inner part for the 
Gamow-Teller transition nearly parallel to that for the Fermi transition. 
The inner part is universal to weak charged-current 
processes and can be absorbed into the modification of the coupling 
constants up to the order of the inverse proton mass $O(1/m_{p})$. 
The resulting $O(\alpha)$ corrections to the differential cross section 
take the form $\left [1+\delta_{\rm out}(E)\right ] 
\left[ (1+\delta_{\rm in}^{\rm F}) 
\langle 1 \rangle^2 +g_A^2 (1+\delta_{\rm in}^{\rm GT}) 
\langle {\mib \sigma} \rangle^2\right]$, where $\langle 1 
\rangle$ and $\langle {\mib \sigma} \rangle$ 
stand for the Fermi and Gamow-Teller matrix elements; 
the outer correction $\delta_{\rm out}$ is positron energy ($E$) 
dependent and takes different functional forms for 
the angular independent and dependent parts. 
All factors are explicitly evaluated. 
\end{abstract} 

\section{Introduction} 

The purpose of this paper is to present a calculation of the radiative 
correction to neutrino-nucleon quasielastic scattering 
(inverse beta decay), 
\begin{eqnarray} 
\bar \nu _{e}+p \longrightarrow e^{+}+ n , 
\label{eq:nubarp} 
\end{eqnarray} 
whose experimental accuracy now approaches the level that makes 
radiative corrections a practically important problem. 
The calculation is similar to that for neutron beta decay, 
which has a long history of no less than forty years 
\cite{kinoshitasirlin}-\cite{marshak}. 

The calculation of radiative corrections to neutron beta decay 
is not quite straightforward. There 
are a number of subtleties. They 
originate from the fact that, at low energies, 
one has to deal with the proton 
and neutron in four-Fermi theory while the loop integral is 
divergent: this divergence is made finite in the electroweak theory, 
but here we must treat quarks and continue the quark calculation 
to hadronic theory at an intermediate energy scale. 

To clarify the issues and the scope of the 
present paper, we start with a 
brief survey of the work done in the past. 
The pioneering work of Kinoshita and Sirlin \cite{kinoshitasirlin}, 
long before the development of renormalizable gauge theories, 
evaluated photon exchange corrections to the neutron beta-decay 
within four-Fermi $V-A$ theory of weak interactions. 
They showed the cancellation of infrared divergences and 
derived the correct velocity dependence of the final 
electron, taming the ultraviolet divergence with 
cut-off theory. 

Repeating the calculation of Kinoshita and Sirlin, 
Berman and Sirlin \cite{berman} speculated that 
the logarithmic divergences are not affected by strong interactions. 
This was later proven by Abers et al.  \cite{abers}. 
The theorem that the logarithmic divergences are universal 
is based on the conserved vector current 
with the use of the current algebra technique \cite{bjorken} \cite{jl}. 
This theorem, however, applies only to the logarithmic divergent part of 
super-allowed Fermi transitions. More precisely speaking, it applies 
only to the purely vector-current contribution in the Fermi 
transitions: it does not apply to the logarithmic divergent part 
arising from the axial current 
which appears on the loop level by interference. 
The divergences due to the axial current 
(which are linear in the axial-vector coupling $g_{A}$) 
in general depend on the model of hadron structure and strong interaction. 
Abers et al. \cite{abers} proposed a model (A$_1$-exchange model) 
to evaluate such contributions.

Sirlin \cite{sirlin} has gone  one step 
further to separate radiative corrections of neutron beta decay 
into outer and inner parts. The outer part corrections are both infrared and 
ultraviolet finite, 
gauge-independent, and contain full electron's velocity dependence, 
thus depending on specific nuclei that undergo beta decay. 
The outer part is not affected by strong interactions. 
The inner part is independent of 
electron's velocity,  infrared finite but 
contains the ultraviolet divergences. 
This separation enables one to absorb the inner part 
into a universal multiplicative correction 
factor of the vector coupling constant that is relevant to all beta decays.

Electroweak gauge theory renders the ultraviolet divergences of the 
inner part finite. 
The calculation \cite{sirlin2,sirlinrev} 
conventionally divides the integration region of the 
virtual gauge bosons into long- and short-distance parts: 
\begin{eqnarray} 
({\rm i}) \hskip0.5cm 0 < \vert k \vert ^{2} < M^{2}, 
\hskip1cm 
({\rm ii}) \hskip0.5cm M^{2} <\vert k \vert ^{2} < \infty , 
\label{eq:region} 
\end{eqnarray} 
where $k$ is the (Wick-rotated Euclidean) momentum of the virtual 
gauge bosons, and the mass scale $M$, introduced by hand, divides 
the low- and high-energy regimes and is supposed 
to lie between the proton-neutron masses ($m_{p}$ and $m_{n}$) and 
the ($W$,$Z$) boson masses, $m_{W}$ and $m_{Z}$. 
Old-fashioned four-Fermi interactions are applied to the proton and 
neutron in region (i), and 
the mass scale $M$ is regarded as the ultraviolet cutoff 
of the QED (i.e., purely photonic) correction. In region (ii), 
electroweak theory is used for quarks and leptons, and 
$M$ is a mass scale that describes the onset of the asymptotic behaviour. 
The concern is  whether the results 
in (i) and (ii) join together smoothly. 
In fact, the theorem mentioned above guarantees that 
the logarithmic terms coming from the pure vector 
current in the super-allowed Fermi 
transitions have the common coefficient in (i) and (ii), and 
the integrals continue 
smoothly; thereby $M$-dependence disappears\footnote{Contrary to 
the logarithmically divergent part that is universal, the constant 
part may in principle receive correction from strong interactions.}. 

Radiative corrections proportional to 
$f_V g_{A}$, on the other hand, requires some intricate treatment. 
(Here $f_{V}=1$ denotes the vector coupling constant parallel to $g_A$; 
we put $f_{V}$ to trace the vector current contributions.) 
One conceivable way  proposed by Sirlin \cite{sirlin2} to 
deal with the axial current contribution is 
to use the free quark model for the short-distance contribution in the 
asymptotic regime, 
and to work explicitly with the proton and neutron for the 
long-distance part, by introducing nucleon's electromagnetic 
and weak form factors to render the logarithmic divergence milder; 
the mass scale $M$ then remains as a lower cut-off 
for the asymptotic regime in the region (ii). This calculation is necessarily 
model-dependent. 
Marciano and Sirlin \cite{marciano} (see also Towner \cite{towner}) 
evaluated the axial current contribution this way, and 
their calculation has widely been employed in $0^+\rightarrow 0^+$ 
beta decay phenomenology.

All the development after Kinoshita and Sirlin, however, is restricted 
to Fermi transitions, whose most important application is 
to $0^{+}\rightarrow 0^{+}$ beta decays; 
little progress has been made regarding 
Gamow-Teller transitions, except for the early studies which showed that 
the outer correction of Sirlin is also applicable 
to Gamow-Teller transitions. 

The full calculation of the radiative correction to (\ref{eq:nubarp}) 
requires not only the treatment of the Fermi transition, but also  
of the Gamow-Teller transition. QED radiative corrections to 
(\ref{eq:nubarp}) 
were already calculated by 
Vogel \cite{vogel2} and Fayans \cite{fayans} within the cut-off 
theory with the point nucleons. From these calculations one can 
extract the outer part whereas the inner part is basically left 
untouched, thus circumventing the problems concerning the axial current 
complications if the inner part is empirically evaluated from 
neutron life time. These calculations, that lack the evaluation of the 
inner part, however, would not completely 
elucidate the structure of radiative corrections in quasi-elastic 
neutrino scattering.

In this paper we attempt to calculate full one-loop radiative corrections to 
(\ref{eq:nubarp}) for both angular independent and dependent parts, 
including those to the Gamow-Teller transition. 
We prove that the coefficient of the logarithmic divergences 
arising from the purely axial vector current are 
not affected by strong interactions, which is 
parallel to the theorem for the Fermi transition\footnote{This proof
is implicitly included in the work of Sirlin \cite{sirlin-sl}, which
discussed the universality of the logarithmic divergence for the
semi-leptonic processes in the electroweak theory. We thank Professor
Sirlin to attract our attention to his work.}. 
The constant term may receive extra contributions from non-conservation of 
the axial current, but our 
observation opens up a way to combine calculations 
in two regions in (\ref{eq:region}) smoothly for the Gamow-Teller part. 

The vector current contribution to the Gamow-Teller transition is 
model dependent, just as much as the axial current part that contributes 
to the Fermi transition. Our treatments will be reciprocal between 
the Fermi and the Gamow-Teller transitions. 

Our observations allow the full calculation of the radiative correction to 
the Gamow-Teller transition 
nearly parallel to that to the Fermi transition, though leaving 
somewhat more room for extra contributions from strong interactions to 
the constant term. 
We present the full expression of the outer 
and inner radiative corrections to 
neutrino-nucleon scattering in the static limit of nucleons. 
The application to other charged current 
processes 
is straightforward. 

\section{Preliminaries}\label{prelimi} 

We use the four-Fermi theory to derive radiative 
corrections in region (i). 
The matrix element of the Born term for the process 
(\ref{eq:nubarp}) is given by 
\begin{eqnarray} 
{\cal M}^{(0)}\equiv \frac{G_{V}}{\sqrt{2}} 
\left [\bar v_{\nu}(p_{\nu})\gamma ^{\lambda }(1-\gamma ^{5}) 
v_{e}(\ell)\right ] 
\left [\bar u_{n}(p_{1})W_{\lambda } 
(p_{1}, p_{2})u_{p}(p_{2})\right ]\ , 
\label{eq:born} 
\end{eqnarray} 
where $G_{V}=G_{F}{\rm cos}\theta _{C}$ with 
the universal Fermi coupling $G_{F}$ and the Cabibbo angle $\theta _{C}$. 
The spinors of the neutron, proton, positron and 
antineutrino are denoted by 
$u_{n}$, $u_{p}$, $v_{e}$, and $v_{\nu}$, respectively, 
with momenta specified in parentheses.

The vertex of the hadronic weak current in general 
depends on nucleon's momenta, thus written as 
$W_{\lambda }(p_{1}, p_{2})$, 
but is well-approximated at low energies by 
\begin{eqnarray} 
W_{\lambda }(p_{1}, p_{2})=  \gamma _{\lambda } (f_V- 
g_{A} \gamma ^{5}), 
\label{eq:weakvertex} 
\end{eqnarray} 
%
%
where $f_V=1$ and  $g_{A}=1.2670$. 
We retain the constant $f_V$ to trace 
the vector current contributions. 
We do not consider the terms of $O(1/m_p)$. 

The differential cross section is given in terms 
of the invariant amplitude ${\cal M} $, 
\begin{eqnarray} 
\frac{d\sigma (\bar \nu _{e}+p \longrightarrow e^{+} + n)} 
{d({\rm cos} \theta )}&=& 
\frac{1}{64 \pi} 
\frac{E\beta }{m_{p}m_{n}E_{\nu}} 
\sum _{{\rm spin}}\vert {\cal M}\vert ^{2} 
\nonumber \\ 
&=& 
\frac{G_{V}^{2}}{2\pi}\: E^{2}\beta 
\left \{ 
A(\beta)+B(\beta)\beta{\rm cos}\theta 
\right \}, 
\label{eq:diffcross} 
\end{eqnarray} 
where $\theta $ is the angle between incident antineutrino 
momentum ${\mib p}_{\nu}$ and positron's, ${\mib \ell }$. 
The spin summation is taken over all external fermions. 
The energy of the positron in the final 
state is denoted by $E=E_{\nu}+m_{p}-m_{n}$, 
$E_{\nu}$ being the incident antineutrino energy; 
$\beta =\sqrt{E^{2}-m_{e} ^{2}}/E$ is the velocity of the positron. 
The forward-backward asymmetry is given by 
\begin{eqnarray} 
\langle {\rm cos}\theta \rangle  =\frac{B(\beta )\beta}{3 A(\beta)}. 
\label{eq:asymmetry} 
\end{eqnarray} 

Upon spin summation for the Born 
amplitude (\ref{eq:born}), we arrive at 
\begin{eqnarray} 
\sum _{{\rm spin}}\vert {\cal M}^{(0)} \vert ^{2} 
=32 G_{V}^{2} m_{n} m_{p} E E_{\nu}\left \{ 
(f_V^{2}+3g_{A}^{2})+(f_V^{2}-g_{A}^{2})\beta {\rm cos}\theta 
\right \}. 
\end{eqnarray} 
In the tree level, 
\begin{eqnarray} 
A(\beta)=A_{0}\equiv f_V^{2}+3g_{A}^{2}  , \qquad 
B(\beta)=B_{0}\equiv f_V^{2}-g_{A}^{2}. 
\label{eq:atob} 
\end{eqnarray} 
We will write $f_V^2=f_V^2\langle 1\rangle^2$, and 
$3g_A^2=g_A^2\langle {\mib \sigma} \rangle^2$ to make the Fermi 
and Gamow-Teller contributions explicit. 

\section{QED corrections} 

The QED radiative corrections in $A(\beta )$  were partly 
obtained by Vogel \cite{vogel2} and Fayans 
\cite{fayans}. These authors, however, did not calculate 
the inner part correction. 
In this paper we compute the full corrections to 
both $A(\beta )$ and $B(\beta )$. 
Some of our formulae presented in this paper are already obtained 
in  \cite{vogel2}  and  \cite{fayans}. 
Nonetheless, since the derivation is not given 
in the above references,  we give sketches of calculations 
so that interested readers can check 
them easily. 

The diagrams we consider are depicted in Figure 1, where 
(v) is the vertex correction, (s) is the self energy correction 
and (b) is bremsstrahlung.  We consider the static limit for 
nucleons, $q^2=(p_1-p_2)^2\ll m_p^2$. 
Our calculation is done in the Feynman gauge 
throughout. 
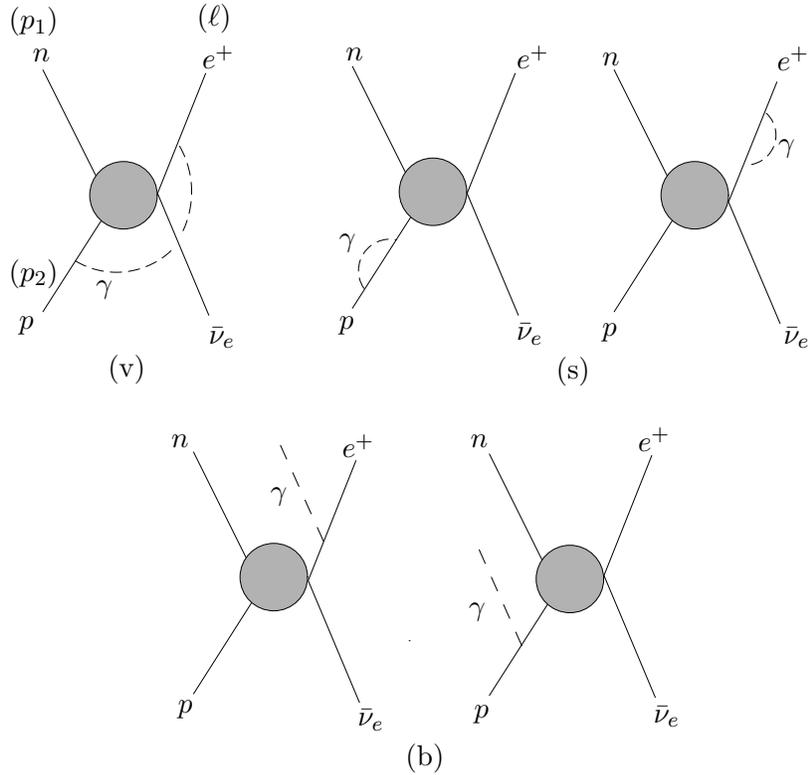
\begin{figure}[htbp] 
\begin{center} 
\input{qed.tex} 
\end{center} 
\caption{QED  corrections: (v) vertex correction, 
(s) self-energy correction, and (b) bremsstrahlung. } 
\end{figure} 

\subsection{Vertex corrections} 

The vertex correction is given by 
\begin{eqnarray} 
{\cal M}^{(v)}&=& 
\frac{i}{\sqrt{2}}G_{V}e^{2}\int \frac{d^{4}k}{(2\pi )^{4}} 
\frac{1}{(k-\ell )^{2}-m_{e}^{2}} 
\frac{1}{(p_{2}-k)^{2}-m_{p}^{2}} 
\frac{1}{k^{2}-\lambda ^{2}} 
\nonumber \\ 
& & \times \bar v_{\nu}\gamma ^{\lambda }(1-\gamma ^{5}) 
\left \{\gamma \cdot (k-\ell )+m_{e}\right \} 
\gamma ^{\mu}v_{e}(\ell) 
\nonumber \\ 
& &\times \bar u_{n}(p_{1})W_{\lambda }(p_{1}, p_{2}-k) 
\left \{\gamma \cdot (p_{2}-k)+m_{p} \right \} 
\gamma _{\mu}u_{p}(p_{2}), 
\label{eq:virtual} 
\end{eqnarray} 
where $\lambda $ is the photon mass to regulate 
the infrared divergence.

Using the identities obtained with the use of the Dirac equation 
for the positron and proton, 
\begin{eqnarray} 
\left \{\gamma \cdot (k-\ell )+m_{e}\right \} 
\gamma ^{\mu}v_{e}(\ell ) 
&=& 
\left \{(k-2\ell )^{\mu}+i\sigma ^{\mu \nu}k_{\nu}\right \} 
v_{e}(\ell ), 
\label{eq:bunkai1} 
\\ 
\left \{ \gamma \cdot (p_{2}-k)+m_{p}\right \} 
\gamma _{\mu}u_{p}(p_{2}) 
&=& 
\left \{(2p_{2}-k)_{\mu}-i\sigma _{\mu \nu}k^{\nu} 
\right \} u_{p}(p_{2}), 
\label{eq:bunkai2} 
\end{eqnarray} 
we decompose (\ref{eq:virtual}) into three parts (see \cite{sirlin}), 
\begin{eqnarray} 
{\cal M}^{(v)}={\cal M}^{(v1)}+{\cal M}^{(v2)}+ 
{\cal M}^{(v3)}. 
\end{eqnarray} 
Here ${\cal M}^{(v1)}$ picks up the product of 
$(k-2\ell )^{\mu}$ in (\ref{eq:bunkai1}) and 
$(2p_{2}-k)_{\mu}$ in (\ref{eq:bunkai2}), and at the same time 
$W_{\lambda }(p_{1}, p_{2}-k)$ is replaced by 
$W_{\lambda }(p_{1}, p_{2})$. 
It  has apparently the same gamma matrix structure 
as the Born term (\ref{eq:born}), and is then written as 
\begin{eqnarray} 
{\cal M}^{(v1)}&=&e^{2} I(\beta ) \times {\cal M}^{(0)}, 
\\ 
I(\beta ) &=&i \int \frac{d^{4}k}{(2\pi )^{4}} 
\frac{(k-2\ell)\cdot (2p_{2}-k)}{ 
\left \{(k-\ell )^{2}-m_{e}^{2}\right \} 
\left \{(p_{2}-k)^{2}-m_{p}^{2}\right \} 
\left \{k^{2}-\lambda ^{2}\right \} 
}. 
\label{eq:sekibuni} 
\end{eqnarray} 
The integral $I(\beta)$ in (\ref{eq:sekibuni}) is given in 
Appendix of \cite{abers}; the real part reads 
\begin{eqnarray} 
I(\beta )+I(\beta )^{*}&=&\frac{1}{8\pi ^{2}}\Bigg [ 
1+{\rm log}\left (\frac{M ^{2}}{m_{e}^{2}}\right )-\frac{2}{\beta} 
{\rm tanh}^{-1}\beta \:{\rm log}\left (\frac{m_{e}^{2}} 
{\lambda ^{2}}\right ) 
\nonumber \\ 
& & 
+\frac{2}{\beta}L\left (\frac{2\beta}{1+\beta}\right )- 
\frac{2}{\beta}({\rm tanh}^{-1}\beta )^{2} 
\Bigg ], 
\end{eqnarray} 
where 
\begin{eqnarray} 
L(z)=\int _{0}^{z}\frac{dt}{t}{\rm log}(1-t). 
\end{eqnarray} 
is the Spence function. 
The ultraviolet cutoff $M $ has been brought about by 
the cut-off of the integral in region (i) of (\ref{eq:region}).

The term ${\cal M}^{(v2)}$ represents the combination of 
$i\sigma ^{\mu \nu}k_{\nu}$ in (\ref{eq:bunkai1}) and 
$(2p_{2}-k)_{\mu}$ in (\ref{eq:bunkai2}),  and 
$W_{\lambda }(p_{1}, p_{2}-k)$ is again replaced with 
$W_{\lambda }(p_{1}, p_{2})$, i.e., 
\begin{eqnarray} 
{\cal M}^{(v2)}&=& 
\frac{i}{\sqrt{2}}G_{V}e^{2}\int \frac{d^{4}k}{(2\pi )^{4}} 
\frac{1}{(k-\ell )^{2}-m_{e}^{2}} 
\frac{1}{(p_{2}-k)^{2}-m_{p}^{2}} 
\frac{1}{k^{2}-\lambda ^{2}} 
\nonumber \\ 
& & \times \bar v_{\nu}\gamma ^{\lambda }(1-\gamma ^{5}) 
i\sigma ^{\mu \nu}k_{\nu} 
v_{e}(\ell) 
\nonumber \\ 
& & \times \bar u_{n}(p_{1})W_{\lambda }(p_{1}, p_{2}) 
(2p_{2}-k)_{\mu}u_{p}(p_{2}). 
\label{eq:ma2} 
\end{eqnarray} 
The gamma matrix structure 
of the nucleon currents in (\ref{eq:ma2}) 
is the same as that in (\ref{eq:born}), although  that of leptons 
is not. The contributions of ${\cal M}^{(v1)}$ and 
${\cal M}^{(v2)}$ are thus evaluated without referring to 
details of the hadronic part of the currents (see Appendix A). 
The radiative correction to the cross section is given as the tree-one-loop 
interference amplitude, which takes the form 
\begin{eqnarray} 
& &\sum _{{\rm spin}} 
\left \{ 
({\cal M}^{(v1)}+{\cal M}^{(v2)}){\cal M}^{(0)*} + 
({\cal M}^{(v1)*}+{\cal M}^{(v2)*}){\cal M}^{(0)} 
\right \} 
\nonumber \\ 
& & 
= 32 G_{V}^{2} m_{n}m_{p} E E_{\nu} 
\Bigg [ 
\left \{ e^{2}\left ( I(\beta )+I(\beta )^{*}\right ) 
+\frac{e^{2}}{4\pi ^{2}} \beta {\rm tanh}^{-1}\beta 
\right \} (f_V^{2}+3g_{A}^{2}) 
\nonumber \\ 
& & 
+\left \{ e^{2}\left ( I(\beta )+I(\beta )^{*}\right ) 
+\frac{e^{2}}{4\pi ^{2}} \frac{1}{\beta }{\rm tanh}^{-1}\beta 
\right \} (f_V^{2}-g_{A}^{2})\beta {\rm cos }\theta 
\Bigg ]. 
\label{eq:311} 
\end{eqnarray}

The remaining terms are collected in ${\cal M}^{(v3)}$, 
which reads 
\begin{eqnarray} 
{\cal M}^{(v3)}&=& 
\frac{i}{\sqrt{2}}G_{V}e^{2}\int \frac{d^{4}k}{(2\pi )^{4}} 
\frac{1}{(k-\ell )^{2}-m_{e}^{2}} 
\frac{1}{(p_{2}-k)^{2}-m_{p}^{2}} 
\frac{1}{k^{2}-\lambda ^{2}} 
\nonumber \\ 
& & \times \bar v_{\nu}\gamma ^{\lambda }(1-\gamma ^{5}) 
\left \{(k-2\ell )^{\mu}+i\sigma ^{\mu \nu}k_{\nu} 
\right \} v_{e}(\ell) 
\nonumber \\ 
& & \times \bar u_{n}(p_{1}) 
R_{\mu \lambda }(p_{1}, p_{2},k)u_{p}(p_{2}), 
\label{eq:ma3} 
\end{eqnarray} 
where 
\begin{eqnarray} 
R_{\mu \lambda}(p_{1}, p_{2}, k)&\equiv & 
\left \{ W_{\lambda }(p_{1}, p_{2}-k)-W_{\lambda }(p_{1}, p_{2}) 
\right \} (2p_{2}-k)_{\mu} 
\nonumber \\  
& &-i W_{\lambda }(p_{1}, p_{2}-k) \sigma _{\mu \nu} k^{\nu} 
\label{eq:rmulambda}
\\ 
&\simeq& -i W_{\lambda }(p_{1}, p_{2}-k) \sigma _{\mu \nu} 
k^{\nu}, 
\end{eqnarray} 
in the approximation of the point nucleons. 
It is only this ${\cal M}^{(v3)}$ term that 
depends on the details of the structure of the 
weak currents. It is clear from the powers of $k$ 
that this term is infrared convergent. 
The explicit use of  the four-Fermi 
interaction (\ref{eq:born}) gives 
an ultraviolet divergence. 
A straightforward  calculation (see Appendix B) shows that 

\begin{eqnarray} 
& & \hskip-1cm \sum _{{\rm spin}}\left \{ 
{\cal M}^{(v3)}{\cal M}^{(0)*}+ 
{\cal M}^{(v3)*}{\cal M}^{(0)} 
\right \} 
\nonumber \\ 
& & = 
32G_{V}^{2}m_{n}m_{p}EE_{\nu} 
\left[ 
\Phi ^{\rm F}(1+\beta {\rm cos}\theta ) 
\langle 1 \rangle ^{2} 
+\Phi ^{\rm GT}(3-\beta {\rm cos}\theta )\cdot 
\frac{1}{3}\langle {\mib \sigma }\rangle ^{2} 
\right], 
\nonumber \\
\label{eq:a3} 
\end{eqnarray} 
where 
\begin{eqnarray} 
\Phi ^{\rm F}&=&\frac{e^{2}}{8\pi ^{2}} 
\left[ f_{V}^{2} \left\{ \frac{3}{2} 
{\rm log}\left (\frac{M^{2}}{m_{p}^{2}} \right ) 
+\frac{3}{4}\right\} 
+{g_{A}}{f_{V}}\left \{ 
\frac{3}{2}{\rm log}\left (\frac{M^{2}}{m_{p}^{2}} \right ) 
+\frac{9}{4} 
\right \} \right ] \ , 
\nonumber \\
\\ 
\Phi ^{\rm GT}&=&\frac{e^{2}}{8\pi ^{2}} \left [ g_{A}^{2}\left\{ 
\frac{3}{2} 
{\rm log}\left (\frac{M^{2}}{m_{p}^{2}} \right ) 
+\frac{7}{4}\right\} 
+{f_{V}}{g_{A}}\left \{ 
\frac{3}{2} 
{\rm log}\left (\frac{M^{2}}{m_{p}^{2}} \right ) 
+\frac{5}{4} \right \} \right ]\ . 
\nonumber \\
\end{eqnarray} 
The superscripts F and GT refer to the contribution to the 
Fermi and Gamow-Teller transitions. Note that the correction from 
${\cal M}^{(v3)}$ is written 
as multiplicative factors on the coupling constants for both Fermi 
and Gamow-Teller parts, without disturbing the original angular 
dependence.

\subsection{Self energy corrections} 

After subtracting the pole term by mass renormalization, 
we are left with 
the wave function renormalization. 
The wave function renormalization constant for 
the fermion of mass $m$  is given by \cite{jauch} 
\begin{eqnarray} 
Z_{2}(m)=1-\frac{e^{2}}{8\pi ^{2}}\left \{ 
\frac{1}{2}{\rm log}\left (\frac{M^{2}}{m^{2}}\right ) +\frac{9}{4} 
-{\rm log}\left (\frac{m^{2}}{\lambda ^{2}}\right ) 
\right \} 
\end{eqnarray} 
for the on-shell renormalization. 
The self energy correction is given by 
\begin{eqnarray} 
{\cal M}^{(s)}&=&\left \{ 
\sqrt{Z_{2}(m_{e})}- 1 + \sqrt{Z_{2}(m_{p})}-1 \right \} 
{\cal M}^{(0)}, 
\label{eq:mc} 
\end{eqnarray} 
and the contribution to the cross section is 
\begin{eqnarray} 
& &\sum _{{\rm spin}}\left ( 
{\cal M}^{(s)}{\cal M}^{(0)*}+{\cal M}^{(s)*}{\cal M}^{(0)} 
\right ) 
\nonumber \\ 
& & 
\hskip1cm =2\left \{ 
\sqrt{Z_{2}(m_{e})}- 1 + \sqrt{Z_{2}(m_{p})}-1 \right \} 
\sum _{\rm spin} \vert {\cal M}^{(0)} \vert ^{2}. 
\label{eq:316} 
\end{eqnarray} 

We see that the $\log M^2$ dependence in (\ref{eq:311}) is cancelled by 
that in (\ref{eq:316}), i.e., the correction from 
${\cal M}^{(v1)}+{\cal M}^{(v2)}+{\cal M}^{(s)}$ is ultraviolet 
finite. The infrared divergence remains that is cancelled after 
the bremsstrahlung contribution ${\cal M}^{(b)}$ is added. 
The combination ${\cal M}^{(v1)}+{\cal M}^{(v2)}+{\cal M}^{(s)}+{\cal 
M}^{(b)}$ 
is identified as the outer part radiative correction of 
Sirlin (up to a numerical constant) for the neutrino-nucleon quasi-elastic 
scattering. It is easy to show that this term is gauge 
independent.  On the other hand, ${\cal M}^{(v3)}$ still contains 
ultraviolet divergence, and it also receives hadronic complications. 
This contribution is taken to be a part of the inner correction.

\subsection{Bremsstrahlung} 

The cross section of single photon emission in Figure 1(b) is given by 
\begin{eqnarray} 
& &\hskip-3cm \sigma (\bar \nu _{e}+p \longrightarrow e^{+} 
+ n + \gamma ) 
=\frac{1}{(2\pi )^{5}}\cdot \frac{1}{8m_{p}m_{n}E_{\nu}}\int 
\frac{d^{3}{\mib \ell '}}{2E'} \int \frac{d^{3}{\mib k}}{2\omega } 
\nonumber \\ 
& & \times \delta(E-E'-\omega) 
\times  \frac{1}{2}\sum _{{\rm spin}}\vert {\cal M}^{(b)} 
\vert ^{2}, 
\label{eq:brems1} 
\end{eqnarray} 
where $ E'=\sqrt{{\mib \ell '}^{2}+m_{e}^{2}}$ and 
$\omega =\sqrt{{\mib k}^{2}+\lambda ^{2}}$. 
In the nucleon static limit, we find 
\begin{eqnarray} 
& &\hskip-1cm \frac{1}{2}\sum _{{\rm spin}}\vert 
{\cal M}^{(b)}\vert ^{2} 
\nonumber \\ 
&=&\frac{64m_{p}m_{n}G_{V}^{2}e^{2}}{[(k+ 
\ell ')^{2}-m_{e}^{2}]}\Bigg \{ 
(f_V^{2}+3g_{A}^{2})E_{\nu}\bigg [ 
E\left ( 
{\mib \ell '}^{2}-\frac{({\mib k}\cdot {\mib \ell '})^{2} 
}{\omega ^{2}}\right ) 
+ (k\cdot \ell ') \omega \bigg ] 
\nonumber \\ 
& & +(f_V^{2}-g_{A}^{2})\bigg [({\mib \ell '}+{\mib k})\cdot 
{\mib p}_{\nu}\left ( 
{\mib \ell '}^{2}-\frac{({\mib k}\cdot {\mib \ell '})^{2} 
}{\omega ^{2}}\right )+(k\cdot \ell ' )({\mib k}\cdot {\mib p}_{\nu}) 
\nonumber \\ 
& &\hskip2cm +(k\cdot \ell )\left ( 
({\mib \ell '}\cdot {\mib p}_{\nu})-\frac{({\mib k}\cdot 
{\mib \ell '})({\mib k}\cdot {\mib p}_{\nu})}{\omega ^{2}} 
\right ) \bigg ]\Bigg \}. 
\label{eq:bremsmatrix} 
\end{eqnarray} 
After integrating over photon phase space, the terms proportional to 
$(f_V^{2}+3g_{A}^{2})$ in 
(\ref{eq:bremsmatrix}) contribute to $A(\beta )$, 
and those proportional to  $(f_V^{2}-g_{A}^{2})$ to 
$B(\beta )$.

Putting (\ref{eq:bremsmatrix}) into (\ref{eq:brems1}) 
and integrating over photon momentum, we obtain 
\begin{eqnarray} 
& &\hskip-2cm \frac{d\sigma (\bar \nu _{e} + p \longrightarrow 
e^{+} + n + \gamma )}{d({\rm cos} \theta )} 
\nonumber \\ 
&=& 
\frac{G_{V}^{2}}{2\pi }E^{2} 
\left (\frac{e^{2}}{8\pi ^{2}}\right ) 
\beta \left \{ 
g^{(b1)}(\beta ) A_{0} + g^{(b2)}(\beta ) 
B_{0} \beta {\rm cos}\theta 
\right \}, 
\end{eqnarray} 
where the function $g^{(b1)}(\beta )$, which was calculated by Vogel 
\cite{vogel2} (partly numerically) and by Fayans \cite{fayans} 
using SCHOONSHIP, is written as 
\begin{eqnarray} 
g^{(b1)}(\beta )&=& 
4{\rm log}\left ( 
\frac{2(E-m_{e})}{\lambda} 
\right )\left (\frac{1}{\beta}{\rm tanh}^{-1}\beta - 1\right ) 
\nonumber \\ 
& &+4\left(\frac{1}{\beta}\tanh^{-1}\beta - 1\right ){\rm log}\left ( 
\frac{2(E+m_{e})}{m_{e}} \right )+ 
\frac{6}{\beta}L\left (\frac{2\beta}{1+\beta}\right ) 
\nonumber \\ 
& & -\frac{6}{\beta }({\rm tanh}^{-1}\beta )^{2}+ 
\frac{7-\beta^2}{2\beta}{\rm tanh}^{-1}\beta + \frac{17}{2}. 
\label{eq:gb1beta} 
\end{eqnarray} 
The other function $g^{(b2)}(\beta )$ is calculated to give 
\begin{eqnarray} 
& &\hskip-1cm g^{(b2)}(\beta )=\frac{4}{\beta ^{2}}+\frac{7}{2}- 
\frac{4\sqrt{1-\beta ^{2}}}{\beta ^{2}} 
\nonumber \\ 
& &-\left ( 
4+\frac{1}{\beta}\right ){\rm tanh}^{-1}\beta 
+ \left ( 
-\frac{1}{2\beta ^{2}}-\frac{3}{2}+\frac{4}{\beta} 
\right )({\rm tanh}^{-1}\beta )^{2} 
\nonumber \\ 
& &-\frac{2}{\beta }L\left ( 
\frac{2\beta}{1+\beta} 
\right )+\frac{8}{\beta}L\left ( 1-\sqrt{ 
\frac{1-\beta}{1+\beta}}\right ) 
\nonumber \\ 
& &+4\left (1-\frac{1}{\beta}{\rm tanh}^{-1}\beta \right ) 
{\rm log}\left \{ 
\frac{\lambda}{2m_{e}}\left (1+\frac{1}{\beta}\right ) 
\frac{\sqrt{1+\beta}+\sqrt{1-\beta}}{\sqrt{1+\beta}- 
\sqrt{1-\beta}} 
\right \}. 
\nonumber \\ 
\label{eq:new1} 
\end{eqnarray} 
The treatment of the infrared divergence for $g^{(b2)}$ is somewhat
subtle.
Steps of the bremsstrahlung calculation are presented in Appendix C. 

\subsection{QED Corrections: Summary} 

We summarise the QED radiative corrections as 
\begin{eqnarray} 
A(\beta) =\langle 1 \rangle^2 
  f_V^2\left [1+\delta_{\rm out}^{\rm F}+{\delta_{\rm in}^{\rm F}}' 
\right ] 
   +\langle {\mib \sigma} \rangle^2 
  g_A^2\left [1+\delta_{\rm out}^{\rm GT}+ 
       {\delta_{\rm in}^{\rm GT}}'\right ]\ , 
\label{eq:Aterm} 
\\ 
B(\beta) = \langle 1\rangle^2 
  f_V^2\left [1+\tilde\delta_{\rm out}^{\rm F}+ 
    {\delta_{\rm in}^{\rm F}}'\right ] 
   -\frac{1}{3}\langle {\mib \sigma} \rangle^2 
  g_A^2\left [1+\tilde\delta_{\rm out}^{\rm GT}+ 
       {\delta_{\rm in}^{\rm GT}}'\right ]\ . 
\end{eqnarray} 
Here 
\begin{eqnarray} 
\delta_{\rm out}^{\rm F}=\delta_{\rm out}^{\rm GT} 
&=&\delta _{\rm out}(E)\cr 
   &=&e^{2}\{ I(\beta )+I(\beta)^{*} \} 
+\frac{e^{2}}{4\pi ^{2}}\beta {\rm tanh}^{-1}\beta \cr 
& & 
+ 2\left \{ 
\sqrt{Z_{2}(m_{e})}- 1 + \sqrt{Z_{2}(m_{p})}-1 \right \}
\nonumber \\
& & 
+\frac{e^{2}}{8\pi ^{2}}\left\{ g^{(b1)}(\beta ) +\frac{3}{4} \right \} 
\label{eq:3/4} 
\\ 
&=& \frac{e^{2}}{8\pi ^{2}}\Bigg [ 
\frac{23}{4}+\frac{3}{2}{\rm log}\left ( 
\frac{m_{p}^{2}}{m_{e}^{2}}\right ) 
+\frac{8}{\beta }L\left ( \frac{2\beta }{1+\beta } \right ) 
\nonumber \\
& & +4{\rm log}\left ( \frac{4\beta ^{2}}{1-\beta ^{2}}\right ) 
\left ( \frac{1}{\beta}{\rm tanh}^{-1}\beta -1  \right ) 
\nonumber \\ 
& & 
-\frac{8}{\beta}({\rm tanh}^{-1}\beta )^{2}+ 
\left ( \frac{7}{2\beta }+\frac{3\beta }{2} 
\right ) {\rm tanh}^{-1}\beta \Bigg ] 
\label{eq:d-out} 
\end{eqnarray} 
is the so-called outer correction, which assembles the contributions from 
${\cal M}^{(v1)}$, ${\cal M}^{(v2)}$, ${\cal M}^{(s)}$, and ${\cal 
M}^{(b)}$. 
This term does not contain 
infrared or ultraviolet divergence, and does not receive complications 
due to hadronic structure and strong interaction. 
All positron-velocity dependences 
are contained in this correction. 
We note that the last term 3/4 in (\ref{eq:3/4}) is added 
(then subtracted from $\delta_{\rm in}$) 
to make the definitions of $\delta_{\rm out}$ 
and $\delta_{\rm in}$ consistent with those of Sirlin for beta 
decay. The outer correction is common to the Fermi and Gamow-Teller 
transitions, and it can be factored out to the order of 
$O(e^{2})$. 

The inner correction ${\delta_{\rm in}}'$ (prime means the correction from QED 
only) arises from ${\cal M}^{(v3)}$. We find 
\begin{eqnarray} 
{\delta_{\rm in}^{\rm F}}' &=& \frac{e^{2}}{8\pi ^{2}}\left [\frac{3}{2} 
\log\left(\frac{M^2}{m_p^2}\right) 
+\frac{g_{A}}{f_{V}}\left \{ \frac{3}{2}{\rm log}\left ( 
\frac{M^{2}}{m_{p}^{2}} \right )+\frac{9}{4}\right \} 
\right]\ , 
\label{eq:in-1-F}\\ 
{\delta_{\rm in}^{\rm GT}}' &=&\frac{e^{2}}{8\pi ^{2}} 
\left[\frac{3}{2} \log\left(\frac{M^2}{m_p^2}\right)+1+ 
\frac{f_{V}}{g_{A}}\left \{ 
\frac{3}{2}{\rm log}\left (\frac{M^{2}}{m_{p}^{2}}\right ) 
+\frac{5}{4}\right \} 
\right]\ . 
\label{eq:in-1-GT} 
\end{eqnarray} 
Here the constant 3/4 is subtracted from 
$\Phi ^{{\rm F}}$ and $\Phi ^{{\rm GT}}$ to 
define ${\delta_{\rm in}^{\rm F}}'$ and 
${\delta_{\rm in}^{\rm GT}}'$. The inner correction is infrared finite, 
but contains ultraviolet divergences, $\log M^2$, which are made finite 
only with electroweak theory. These terms do not contain positron-velocity 
dependent factors, and thus absorbed into 
the vector and axial-vector coupling constants. They also agree with 
the inner correction for beta decay. The inner correction for the Fermi 
transition has been identified in \cite{sirlin}.  
The first logarithmic divergent 
parts, both in  ${\delta_{\rm in}^{\rm F}}'$ and ${\delta_{\rm in}^{\rm 
GT}}'$, do not receive corrections (see Sect. \ref{sec:universality}). 
However, the constant terms 
may, in principle, receive corrections from strong interactions. 
We shall argue that effects of the hadron structure are essential for 
terms proportional to $g_{A}/f_{V}$ in (\ref{eq:in-1-F}) and 
$f_{V}/g_{A}$ in (\ref{eq:in-1-GT}),  including 
the coefficients of logarithmic divergences.

For the correction of the angular dependent part, we have 
\begin{eqnarray} 
\tilde\delta_{\rm out}^{\rm F}= 
\tilde\delta_{\rm out}^{\rm GT}&=& 
\tilde \delta _{\rm out}(E)\cr 
   &=&e^{2}\{ I(\beta )+I(\beta)^{*} \} 
+\frac{e^{2}}{4\pi ^{2}}{1\over \beta} {\rm tanh}^{-1}\beta \cr 
& & 
+ 2\left \{ 
\sqrt{Z_{2}(m_{e})}- 1 + \sqrt{Z_{2}(m_{p})}-1 \right \} 
+\frac{e^{2}}{8\pi ^{2}}\left\{ g^{(b2)}(\beta ) +\frac{3}{4} \right \} 
\cr 
&=& \frac{e^{2}}{8\pi ^{2}}\Bigg [ 
\frac{3}{4} + \frac{3}{2}{\rm log}\left ( 
\frac{m_{p}^{2}}{m_{e}^{2}}\right )  
\nonumber \\
& & 
+\frac{8}{\beta } L\left ( 
1-\sqrt{\frac{1-\beta }{1+\beta }}\right ) 
+ 
\frac{4}{\beta ^{2}}-\frac{4\sqrt{1-\beta ^{2}}}{\beta ^{2}} 
\nonumber \\ 
& & +4 \left (1-\frac{1}{\beta }{\rm tanh}^{-1}\beta \right ) {\rm log} 
\left ( \frac{1}{2}\left (1+\frac{1}{\beta }\right ) 
\frac{\sqrt{1+\beta }+\sqrt{1-\beta }}{\sqrt{1+\beta }-\sqrt{1-\beta }} 
\right ) 
\nonumber \\ 
& & + \left ( \frac{1}{\beta }-4\right ){\rm tanh}^{-1}\beta 
+ \left (\frac{2}{\beta }-\frac{3}{2}-\frac{1}{2\beta ^{2}}\right ) 
({\rm tanh}^{-1}\beta )^{2}\Bigg ]. 
\label{eq:d-out2} 
\end{eqnarray} 
The inner correction ${\delta_{\rm in}^{\rm F}}'$ and 
${\delta_{\rm in}^{\rm GT}}'$ 
are, as evident from (\ref{eq:a3}), 
identical with those that appear in the angular independent correction 
(\ref{eq:Aterm}); hence we have not attached tildes. 

\section{Electroweak corrections and the continuation to low energy} 
\label{sec:electroweak}

The short distance correction from the integration region (ii) 
in (\ref{eq:region}) is evaluated using electroweak theory \cite{sirlin2}. 
For the application of electroweak theory, we must deal with 
quarks, so that the continuation of quark theory to low-energy 
effective theory for nucleons is essential. 
When we consider corrections 
relative to muon decay, we only need to consider the photon and 
$Z$ exchange diagrams shown in Figure 2.

\begin{figure}[htbp] 
\begin{center} 
\input{wz.tex} 
\end{center} 
\caption{Photon and Z-boson exchange diagrams} 
\label{fig:gammaZ} 
\end{figure}
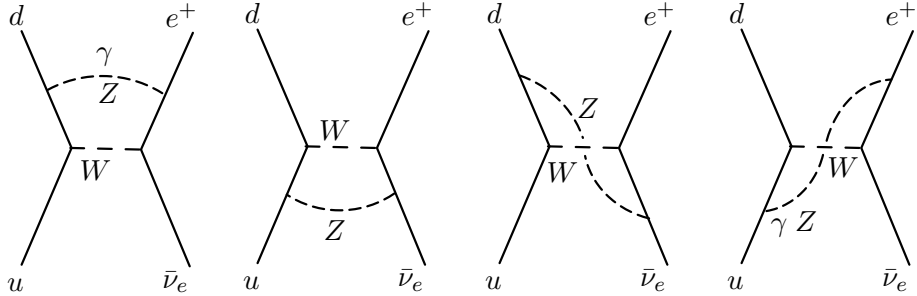 

For a short distance calculation, we can neglect mass and 
momenta of the external fermions. 
This calculation was done in \cite{sirlin2}. 
A calculation for the photon exchange yields, 
\begin{eqnarray} 
\hat {\cal M}^{(\gamma W)}= 
\frac{e^{2}}{16\pi ^{2}}\left ( 
\frac{3}{2}+3\bar Q 
\right ){\rm log}\left (\frac{m_{W}^{2}}{ 
M^{2}}\right ){\hat{\cal M}}^{(0)}\ , 
\label{eq:gammaw} 
\end{eqnarray} 
where  ${\hat {\cal M}}^{(0)}$ is the 
tree amplitude for neutrino scattering off the assembly of free 
quarks, and ${\bar Q}=1/6$ is the mean 
charge of the up and down quarks. Note that 
the second term (proportional to $\bar Q$) originates from the 
product of two antisymmetric tensors 
$\varepsilon ^{\lambda \mu \nu \rho}$ of 
leptonic and quark gamma matrices. 
Similarly for the $Z$-boson exchange, we obtain 
\begin{eqnarray} 
{\hat{\cal M}}^{(ZW)}=\frac{e^{2}}{16\pi ^{2}} 
\left ( 3{\bar Q}+\frac{5}{2{\rm tan}^{4}\theta _{W}}\right ) 
{\rm log}\left ( \frac{m_{Z}^{2}}{m_{W}^{2}} \right ) 
{\hat {\cal M}}^{(0)}\ . 
\label{eq:zw} 
\end{eqnarray} 
The first term is again due to the product of two antisymmetric tensors. 
The second term is all the rest.

In order to connect the quark-level amplitudes with hadronic ones, 
we assume that the ratio of the tree and 
loop amplitudes for the neutrino quark scattering 
is the same as that for the neutrino nucleon scattering 
\cite{sirlin2}. This is justified at least for the logarithmic 
divergent part of the correction for the Fermi transition by 
current conservation. In the next section we justify 
this for the  logarithmic 
divergent part of the correction for the Gamow-Teller transition 
in the presence of partial conservation of the axial current. 
With this prescription we write the radiative correction 
to neutrino nucleon scattering at the short distance as, 
\begin{eqnarray} 
\left \{ {\hat {\cal M}}^{(0)}\left ( {\hat {\cal M}}^{(\gamma W)*}+ 
{\hat {\cal M}}^{(ZW)*}\right )+ 
{\hat {\cal M}}^{(0)*}\left ( {\hat {\cal M}}^{(\gamma W)} 
+{\hat {\cal M}}^{(ZW)}\right ) 
\right \}\frac{\vert {\cal M}^{(0)}\vert ^{2}}{\vert {\hat 
{\cal M}}^{(0)}\vert ^{2}} 
\nonumber \\ 
= 
\frac{e^{2}}{8\pi ^{2}}\left \{ 
\frac{3}{2} {\rm log}\left (\frac{m_{W}^{2}}{M^{2}} 
\right ) +3\bar Q {\rm log}\left ( 
\frac{m_{Z}^{2}}{M^{2}} 
\right ) 
+\frac{5}{2{\rm tan}^{4}\theta _{W}}{\rm log}\left ( 
\frac{m_{Z}^{2}}{m_{W}^{2}}\right ) 
\right \} 
\vert {\cal M}^{(0)}\vert ^{2}. 
\nonumber \\
\label{eq:wa} 
\end{eqnarray} 
We observe that the $M$ dependence (upper cutoff) that appears in 
the long-distance radiative correction 
(\ref{eq:in-1-F}) is cancelled by the first term in the brackets 
in (\ref{eq:wa}), 
which renders the ultraviolet divergence in the QED radiative correction 
finite 
\cite{sirlin2}. 
Rigorously speaking, 
this universality applies only to the logarithmically divergent part, and 
the 
constant terms might receive extra contributions.
Here we simply take the calculation with the point nucleon for the 
constant term.

Note that the weak coupling constant is determined by the rate of 
muon decay, where the 
same box-type Feynman integrals appears. 
The formula used to determine $G_F$ from muon lifetime 
\begin{eqnarray} 
\frac{1}{\tau _{\mu}}= 
\frac{{G}_{F}^{2}m_{\mu}^{5}}{192 \pi ^{3}} 
\left ( 1-\frac{8m_{e}^{2}}{m_{\mu}^{2}}\right ) 
\left \{ 
1+\frac{3m_{\mu}^{2}}{5m_{W}^{2}}+\frac{e^{2}}{8\pi ^{2}} 
\left [ 
\frac{25}{4}-\pi ^{2} 
\right ] 
\right \} 
\label{eq:conventional} 
\end{eqnarray} 
means that 
the effects of the box-type diagrams in muon decay 
is absorbed in $G_{F}$. 
Thus, when we calculate 
radiative corrections in terms of $G_{F}$, we should subtract 
those  included in $G_{F}$: the relevant contributions for 
muon decay are obtained by putting 
$\bar Q=-1/2$ in (\ref{eq:wa}): 
\begin{eqnarray} 
\frac{e^{2}}{8\pi ^{2}}\left \{ 
\frac{3}{2} {\rm log}\left (\frac{m_{W}^{2}}{M^{2}} 
\right ) -\frac{3}{2} {\rm log}\left ( 
\frac{m_{Z}^{2}}{M^{2}} 
\right ) 
+\frac{5}{2{\rm tan}^{4}\theta _{W}}{\rm log}\left ( 
\frac{m_{Z}^{2}}{m_{W}^{2}}\right ) 
\right \} 
\vert {\cal M}^{(0)}\vert ^{2} 
\nonumber \\ 
= 
\frac{e^{2}}{8\pi ^{2}}\left ( 
-\frac{3}{2} +\frac{5}{2{\rm tan}^{4}\theta _{W}}\right ) 
{\rm log}\left ( 
\frac{m_{Z}^{2}}{m_{W}^{2}}\right ) 
\vert {\cal M}^{(0)}\vert ^{2}. 
\label{eq:muon} 
\end{eqnarray}


Adding (\ref{eq:wa}) to the QED corrections 
${\delta _{{\rm in}}^{\rm F}}'$ 
and subtracting (\ref{eq:muon}), 
the Fermi part inner correction of (\ref{eq:in-1-F}) is 
\begin{eqnarray} 
\delta _{\rm in}^{\rm F}&\equiv & 
{\delta _{{\rm in}}^{\rm F}}'+ 
\frac{e^{2}}{8\pi ^{2}}\left[ \frac{3}{2} 
\log\left(\frac{m_{W}^{2}}{M^{2}}\right) 
+3{\bar Q} \log\left(\frac{m_{Z}^{2}}{M^{2}}\right) 
+\frac{5}{2{\rm tan}^{4}\theta _{W}}{\rm log}\left ( 
\frac{m_{Z}^{2}}{m_{W}^{2}}\right )\right ] 
\nonumber \\ 
& &-\frac{e^{2}}{8\pi ^{2}} 
\left (-\frac{3}{2}+\frac{5}{2{\rm tan}^{4}\theta _{W}} 
\right ) 
{\rm log}\left ( 
\frac{m_{Z}^{2}}{m_{W}^{2}}\right ) 
\nonumber \\ 
&=& 
\frac{e^{2}}{8\pi ^{2}}\left[ \frac{3}{2} 
\log\left(\frac{m_{Z}^{2}}{m_{p}^{2}}\right) 
+3{\bar Q} \log\left(\frac{m_{Z}^{2}}{M^{2}}\right) 
+C_{}^{\rm F}\right ]\ , 
\label{eq:summe} 
\end{eqnarray} 
where the terms proportional to $g_{A}/f_{V}$ are collected in $C_{}^{\rm 
F}$, 
\begin{eqnarray} 
C_{}^{\rm F}= \frac{g_{A}}{f_{V}} 
\left \{ 
\frac{3}{2}{\rm log}\left ( 
\frac{M^{2}}{m_{p}^{2}}\right )+\frac{9}{4} 
\right \} .
\end{eqnarray} 

As noted above, the $M$-dependence in the 
first logarithm in (\ref{eq:summe}) 
is cancelled by the term $\frac{{\displaystyle 3}}{{\displaystyle 2}} 
{\rm log}\left ( 
M^{2}/m_{p}^{2}\right )$ in ${\delta _{{\rm in}}^{\rm F}}'$. 
The other $M$ dependent  term 
$(3g_{A}/2f_{V}){\rm log}(M^{2}/m_{p}^{2})$ in ${\delta_{{\rm in}}^{\rm 
F}}'$ 
that arises from the vector-axial-vector interference, 
however, is hadron structure dependent, which is manifest by the fact that 
it is not cancelled by the electroweak  counterpart 
$3\bar Q {\rm log}(m_{Z}^{2}/M^{2})$: 
the coefficients of the two logarithmic terms agree only for an 
unrealistic value 
$\bar Q=g_{A}/2f_{V}$. For the evaluation of 
these contributions, one  separates long- and short-distance 
contributions \cite{sirlin2,marciano}, leaving the 
$M$ dependence that appear in 
(\ref{eq:summe})  as the lower cut-off of the 
short-distance integral, whereas estimating 
the long distance contribution $C^{\rm F}$ 
by rendering its logarithmic dependence milder with 
nucleon structure taken into account \cite{marciano}.

We can proceed in a parallel manner for the Gamow-Teller transition. 
The basic observation is the universality of the logarithmic 
divergence that also applies to the Gamow-Teller transition for 
the axial-axial contribution, as shown in the next section. 
With this universality we can identify the QED correction 
given in terms of nucleon matrix elements in (\ref{eq:in-1-GT}) 
with that of free quark theory, (\ref{eq:gammaw}), and the match 
of the coefficients means the cancellation of the $M$ dependences 
between the two expressions. 
We cannot prove the universality for the constant terms and those 
proportional to $f_{V}/g_{A}$. 
The corrections for the Gamow-Teller transition 
may have a more room of extra contributions from 
strong interaction than for the Fermi transition, unless chiral 
symmetry is exact.  Again, we simply assume the calculation of the point 
nucleons as was done for the Fermi transition except for those 
proportional to $f_{V}/g_{A}$. We obtain for the Gamow-Teller 
transition, 
\begin{eqnarray} 
\delta _{\rm in}^{\rm GT}&=& 
\frac{e^{2}}{8\pi ^{2}}\left[\frac{3}{2} 
\log\left(\frac{m_{Z}^{2}}{m_p^2}\right) 
+1+ 
3{\bar Q} 
\log\left(\frac{m_{Z}^2}{M^{2}}\right) 
+C_{}^{\rm GT} 
\right]\ , 
\label{eq:d-in-gt} 
\end{eqnarray} 
where the terms proportional to $f_{V}/g_{A}$ collected in 
$C_{}^{\rm GT}$ are 
\begin{eqnarray} 
C_{}^{\rm GT}= \frac{f_{V}}{g_{A}} 
\left \{ 
\frac{3}{2}{\rm log}\left ( 
\frac{M^{2}}{m_{p}^{2}} \right ) +\frac{5}{4} 
\right \} 
\label{eq:1258} 
\end{eqnarray} 
for point nucleons. 

For the vector-axial-vector interference term, 
we encounter exactly the same problem as for the Fermi transition. 
The two logarithmic terms, 
$3\bar Q {\rm log}(m_{Z}^{2}/M^{2})$ in (\ref{eq:d-in-gt}) and 
$(3f_{V}/2g_{A}) 
{\rm log}(M^{2}/m_{p}^{2})$ in (\ref{eq:1258}), 
match only when $\bar Q=f_{V}/2g_{A}$, 
another unrealistic value. 
So, we take the same prescription carried out for the 
Fermi transition by Marciano and Sirlin \cite{marciano}, 
by including  the long-distance part correction introducing nucleon 
form factors. We present the calculation of $C_{}^{\rm GT}$ 
in Sect.\ref{sec:cpara}. 

\section{Universality of the logarithmic divergent part for the 
Gamow-Teller transition} 
\label{sec:universality}

Abers et al. \cite{abers} gave a proof that the logarithmic 
divergences 
$\frac{{\displaystyle 3}}{{\displaystyle 2}}{\rm log}(M^{2}/m_{p}^{2})$ 
that appear in the Fermi part, i.e., 
$\delta _{\rm in}^{\rm F'}$, 
is universal in that their coefficients are model-independent regarding 
the structure of hadrons: the coefficient of the logarithmic 
divergences does not 
depend on whether one deals with hadronic or quark current, in so far 
as these currents satisfy the generic commutation relations. 
Here, we argue that the same statement applies to the Gamow-Teller 
counterpart in 
$\delta  _{\rm in}^{\rm GT'}$. 
The currents and the current commutation relations are given in Appendix 
\ref{app:ca}.  We note that the proof described here is implicitly
included in the work of Sirlin \cite{sirlin-sl}, which shows that
the logarithmic divergences that appear in semi-leptonic processes
are universal in the electroweak theory. We show explicitly the
universality of the divergences between the four Fermi theory and
the electroweak theory for the axial-vector correction to the
Gamow-Teller part.

We introduce two types of Green's functions: 
\begin{eqnarray} 
T_{\lambda \mu}(k, p_{1}, p_{2})&=&i\int d^{4}x e^{ik\cdot x} 
\langle p_{1} \vert T \left ( 
t_{\lambda }(0)j^{{\rm e.m.}}_{\mu }(x) 
\right )\vert p_{2}\rangle , 
\label{eq:green1} 
\\ 
T_{\lambda \mu \nu}(k, q, p_{1}, p_{2})&=&\int d^{4}x e^{iq \cdot x} 
\int d^{4}y e^{ik\cdot y}\langle p_{1} \vert T\left (t_{\lambda }(x) 
j^{{\rm e.m.}}_{\mu }(y)j^{{\rm e.m.}}_{\nu}(0) 
\right )\vert p_{2}\rangle 
\nonumber \\ 
& & 
-\delta T_{\lambda \mu \nu}\ ,
\label{eq:green2} 
\end{eqnarray} 
where $j_{\mu}^{{\rm e.m.}}(x)$ is the  electromagnetic 
current and    $t_{\lambda }(x)$ 
is the $V-A$ weak current
\begin{eqnarray} 
t_{\lambda }(x)= V_{\lambda }(x)- A_{\lambda }(x). 
\label{eq:weakcurrent} 
\end{eqnarray} 
The term $-\delta T_{\lambda \mu \nu}$ in (\ref{eq:green2}) 
subtracts the pole term at $p_{2}^{2}=m_{p}^{2}$ in 
$T_{\lambda \mu \nu}(k, p_{1}, p_{2})$, 
which corresponds to mass renormalization.
The considerations in what 
follows does not depend whether the weak current 
(\ref{eq:weakcurrent}) is expressed in terms of 
quark  or nucleon fields,  
because our discussion is based solely on the 
current-current commutation relation between vector and axial-vecrtor 
currents.

With conservation of the electromagnetic current and 
current algebra applied to (\ref{eq:green1}) and (\ref{eq:green2}), we 
obtain identities: 
\begin{eqnarray} 
k^{\mu }T_{\lambda \mu}(k, p_{1}, p_{2})&=&\langle p_{1} \vert 
t_{\lambda }(0)\vert p_{2}\rangle , 
\label{eq:ward1} 
\\ 
k^{\lambda }T_{\lambda \mu}(k, p_{1}, p_{2}) 
&=&\langle p_{1} \vert t_{\mu }(0) 
\vert p_{2}\rangle +M_{\mu}-(p_{1}-p_{2})^{\lambda}T_{\lambda \mu}( 
k, p_{1}, p_{2}), 
\nonumber \\
\label{eq:ward2} 
\\ 
k^{\mu }T_{\lambda \mu \nu}(k. q, p_{1}, p_{2})&=& 
-T_{\lambda \nu}(p_{2}-p_{1}-k-q, p_{1}, p_{2})- 
k^{\mu }\delta T_{\lambda \mu \nu}, 
\label{eq:ward3} 
\\ 
q^{\lambda }T_{\lambda \mu \nu}(k, q, p_{1}, p_{2}) &=&
T_{\mu \nu}(p_{2}-p_{1}-k-q, p_{1}, p_{2})+T_{\nu \mu}(k, p_{1}, p_{2}) 
\nonumber \\ 
& &+M_{\mu \nu}-q^{\lambda }\delta T_{\lambda \mu \nu}, 
\label{eq:axialnonconservation} 
\end{eqnarray} 
where 
\begin{eqnarray} 
M_{\mu}&=& - \int d^{4}x e^{i(p_{2}-p_{1}-k) \cdot x}\langle p_{1} \vert 
T( 
\partial \cdot A(x) j_{\mu}^{{\rm e.m.}}(0)\vert p_{2}\rangle , 
\end{eqnarray}
and 
\begin{eqnarray}
M_{\mu \nu}&=&-i \int d^{4}x  e^{iq\cdot x}\int d^{4}y e^{ik\cdot y} 
\langle p_{1} \vert T\left ( 
\partial \cdot A(x) j^{{\rm e.m.}}_{\mu }(y)j^{{\rm e.m. }}_{\nu }(0) 
\right ) \vert p_{2}\rangle 
\nonumber \\
\end{eqnarray} 
arise in (\ref{eq:ward2}) and (\ref{eq:axialnonconservation}), 
as a result of non-conservation of the axial current. 
If we can set 
$p_{2}-p_{1}\approx 0$, then (\ref{eq:ward2}) is simplified 
as 
\begin{eqnarray} 
k^{\lambda }T_{\lambda \mu}(k, p_{1}, p_{2})&=&\langle p_{1} \vert 
t_{\mu }(0) 
\vert p_{2}\rangle +M_{\mu}. 
\label{eq:ward22} 
\end{eqnarray} 

By differentiating (\ref{eq:axialnonconservation}) with respect to 
$q^{\lambda }$ and setting $q=p_{2}-p_{1}$, we get the  identity 
\begin{eqnarray} 
& &T_{\lambda \mu \nu}(k, p_{2}-p_{1}, p_{1}, p_{2})+ 
q^{\rho}\frac{\partial}{\partial q^{\lambda}} 
T_{\rho \mu \nu}(k, q, p_{1}, p_{2})\Bigg \vert _{q=p_{2}-p_{1}} 
\nonumber \\ 
& &\hskip1cm =\frac{\partial }{\partial k^{\lambda }}T_{\mu \nu} 
(-k, p_{1}, p_{2})+\frac{\partial }{\partial q^{\lambda }} 
(M_{\mu \nu}-q^{\rho}\delta T_{\rho \mu \nu})\Bigg \vert _{q=p_{2}-p_{1}}. 
\end{eqnarray} 
Neglecting again $q=p_{2}-p_{1}$, we write 
\begin{eqnarray} 
T_{\lambda \mu \nu}(k, p_{2}-p_{1}, p_{1}, p_{2}) 
=\frac{\partial }{\partial k^{\lambda }}T_{\mu \nu} 
(-k, p_{1}, p_{2})+\frac{\partial }{\partial q^{\lambda }} 
(M_{\mu \nu}-q^{\rho}\delta T_{\rho \mu \nu})
\Bigg \vert _{q=p_{2}-p_{1}}. 
\nonumber \\
\label{eq:c12} 
\end{eqnarray} 

We redefine ${\cal M}^{(v)}$ of Figure 1, including the effect of 
strong interaction in terms of the Green's functions (\ref{eq:green1}) 
and (\ref{eq:green2}): 
\begin{eqnarray} 
{\cal M}^{(v)}&=&-\frac{G_{V}}{\sqrt{2}}e^{2}\int \frac{d^{4}k}{(2\pi )^{4}} 
\frac{1}{k^{2}-\lambda ^{2}} 
\bar v_{\nu}\gamma ^{\lambda }(1-\gamma ^{5})S_{F}(k-\ell )\gamma ^{\mu} 
v_{e}(\ell ) 
T_{\lambda \mu}(k, p_{1}, p_{2}) 
\nonumber \\ 
&=& 
-\frac{G_{V}}{\sqrt{2}}e^{2}\int \frac{d^{4}k}{(2\pi )^{4}} 
\frac{1}{k^{2}-\lambda ^{2}}\frac{i}{(k-\ell )^{2}-m_{e}^{2}}T_{\lambda \mu} 
(k, p_{1}, p_{2}) 
\nonumber \\ 
& & \times \bar v_{\nu}\bigg ( 
k^{\lambda }\gamma ^{\mu}+k^{\mu}\gamma ^{\lambda }-g^{\lambda \mu}\gamma 
\cdot k -i\varepsilon ^{\lambda \rho  \mu \sigma  }\gamma _{\sigma} 
\gamma ^{5}k_{\rho }-2\ell ^{\mu}\gamma ^{\lambda } 
\Bigg )(1-\gamma ^{5})v_{e}(\ell ) 
\nonumber \\ 
\label{eq:c11} 
\end{eqnarray} 
after some manipulation of gamma matrices in the leptonic part. The 
first and second terms in the brackets of (\ref{eq:c11}) 
can  be expressed  by using (\ref{eq:ward1}) and (\ref{eq:ward22}) 
in terms of the single-current matrix element. 

The third term requires a manipulation of 
$g^{\lambda \mu}T_{\lambda \mu}(k, p_{1}, p_{2})$. 
It is known 
\cite{bjorken,jl} that the large energy limit of the Green's function 
(\ref{eq:green1}) may be expressed in terms of the equal-time (ET) 
current commutator 
\begin{eqnarray} 
{\rm lim}_{k^{0}\rightarrow +\infty }T_{\lambda \mu } 
(k, p_{1}, p_{2}) = 
\frac{1}{k^{0}}\int d^{3}{\mib x} e^{-i{\mib k}\cdot {\mib x}} 
\langle p_{1}  \vert \left [ 
t_{\lambda }(0),  j^{{\rm e.m.}}_{\mu}(x) 
\right ]_{{\small {\rm ET}}} \vert p_{2}\rangle . 
\label{eq:bjl} 
\end{eqnarray} 
The current algebra relation, used in  (\ref{eq:bjl}), 
\begin{eqnarray} 
g^{\lambda \mu} 
\left [t_{\lambda }(0), j^{{\rm e.m.}}_{\mu}(x) 
\right ]_{{\rm ET}}=-2t_{0}(x)\delta ^{3}({\mib x}) 
\end{eqnarray} 
determines 
the high energy behaviour of 
$g^{\lambda \mu}T_{\lambda \mu}(k, p_{1}, p_{2})$, which reads 
\begin{eqnarray} 
{\rm lim}_{k^{0}\rightarrow +\infty }T_{\lambda \mu } 
(k, p_{1}, p_{2}) g^{\lambda \mu} 
=-\frac{2}{k^{0}}\langle p_{1} \vert t_{0}(0)\vert p_{2} \rangle . 
\end{eqnarray} 
This is  non-covariant looking. If we are interested only in 
the leading term in $k^2$, we can follow the trick of \cite{abers} 
to rewrite it in a covariant form: 
\begin{eqnarray} 
{\rm lim}_{k^{2}\rightarrow +\infty } 
T_{\lambda \mu }(k, p_{1}, p_{2}) g^{\lambda \mu}\longrightarrow 
-\frac{2k^{\mu}}{k^{2}-2k\cdot p_{2}}\langle p_{1} \vert t_{\mu}(0)\vert 
p_{2} \rangle . 
\label{eq:highenergy1} 
\end{eqnarray}

The evaluation of the fourth term in (\ref{eq:c11}) requires knowledge 
beyond generic current commutation relations. 
Writing explicitly the current in terms of 
nucleons or quarks (see Appendix \ref{app:ca}), we obtain 
\begin{eqnarray} 
-i\varepsilon ^{\lambda \rho \mu \sigma} k_{\rho } 
\left [ 
t_{\lambda }(0), j^{{\rm e.m.}}_{\mu }(x) 
\right ]_{{\rm ET}}=4\bar Q \left \{ 
k_{0}  t ^{\sigma }(0)-g^{\sigma }_{0}k \cdot  t (0) 
\right \} \delta ^{3}({\mib x}), 
\label{fgaterm} 
\end{eqnarray} 
where 
$\bar Q$ is the mean electric charge of the fundamental  doublet, 
i.e., $\bar Q=1/2$ for proton and neutron, or $+1/6$ for quarks. 
The presence of $\bar Q$ means that 
this current commutator is not model-independent.
Note also that the commutators used to derive this expression
are those beyond ordinary current algebra; they are determined
only with the aid of models of hadrons (see Appendix D). 
We see that the antisymmetric tensor in the 
l.h.s. of (\ref{fgaterm}) converts the vector (axial-vector)
currents into the axial-vector (vector) currents.
Therefore, the $g_A$ dependence appears for the vector current
contributions upon taking the hadronic matrix element.  
 This contrasts to the situation for the 
first three terms in (\ref{eq:c11}), which are determined by 
current algebra and conservation laws; the effect of strong
interaction for the axial-vector current is properly absorbed
in $g_A$ after taking the matrix element with nucleons.

We are interested only in the leading high energy behavior and 
again write (\ref{fgaterm}) in a covariant form: 
\begin{eqnarray} 
& &-i\: {\rm lim}_{k^{2}\rightarrow +\infty} 
\varepsilon ^{\lambda \rho \mu \sigma }k_{\rho } 
T_{\lambda \mu }(k, p_{1}, p_{2}) 
\nonumber \\ 
& & \hskip2cm \longrightarrow 
4\bar Q \left ( 
g^{\kappa \sigma }-\frac{k^{\kappa }k^{\sigma }}{k^{2}} 
\right )\langle p_{1} \vert  t_{\kappa }(0) \vert p_{2}\rangle 
\frac{k^{2}}{k^{2}-2k\cdot p_{2}}. 
\label{eq:highenergy2} 
\end{eqnarray} 

The fifth term in the brackets of (\ref{eq:c11}) does not contribute 
to its divergent part. Thus, the divergent part of (\ref{eq:c11}) 
is evaluated as, 
\begin{eqnarray} 
{\cal M}^{(v)}\Bigg \vert _{\rm divergent}
&=& 
-\frac{G_{V}}{\sqrt{2}}e^{2}\int \frac{d^{4}k}{(2\pi )^{4}} 
\frac{1}{k^{2}-\lambda ^{2}}\frac{i}{(k-\ell )^{2}-m_{e}^{2}} 
\bar v_{\nu}\gamma ^{\lambda }(1-\gamma ^{5})v_{e}(\ell ) 
\nonumber \\ 
& & \times 
\Bigg \{ 2\langle p_{1}\vert t_{\lambda }(0) \vert p_{2}\rangle +M_{\lambda} 
+\frac{2k^{\mu}k_{\lambda}}{k^{2}-2k\cdot p_{2}} 
\langle p_{1}\vert t_{\mu }(0) \vert p_{2}\rangle 
\nonumber \\ 
& & 
+4\bar Q \left ( 
g^{\kappa }_{\lambda}-\frac{k^{\kappa }k_{\lambda }}{k^{2}} 
\right )\langle p_{1}\vert  t_{\kappa }(0) \vert p_{2}\rangle 
\frac{k^{2}}{k^{2}-2k\cdot p_{2}} 
\Bigg \} 
\nonumber \\ 
&\sim & 
-\frac{G_{V}}{\sqrt{2}} 
e^{2}\int \frac{d^{4}k}{(2\pi )^{4}} 
\frac{1}{k^{2}-\lambda ^{2}}\frac{i}{(k-\ell )^{2}-m_{e}^{2}} 
\bar v_{\nu}\gamma ^{\lambda }(1-\gamma ^{5})v_{e}(\ell ) 
\nonumber \\ 
& &\times 
\Bigg \{ 
\left (2+\frac{1}{2}\right ) 
\langle p_{1}\vert  t_{\lambda }(0) \vert p_{2}\rangle 
+M_{\lambda} 
\nonumber \\
& & 
\hskip0.5cm 
+3\bar Q \langle p_{1}\vert   t_{\lambda }(0) \vert p_{2}\rangle 
\frac{k^{2}}{k^{2}-2k\cdot p_{2}} 
\Bigg \} 
\label{eq:mlambda}
\\ 
&\sim &\frac{e^{2}}{16\pi ^{2}}{\rm log }M^{2} \left ( 
\frac{5}{2}{\cal M}^{(0)}+3\bar Q  {\cal M}^{(0)} 
\right )\ , 
\label{eq:c23} 
\end{eqnarray} 
where 
\begin{eqnarray} 
{\cal M}^{(0)}=\frac{G_{V}}{\sqrt{2}} 
\bar v_{\nu}\gamma ^{\lambda }(1-\gamma ^{5})v_{e}(\ell ) 
\langle p_{1}\vert  t_{\lambda }(0) \vert p_{2}\rangle 
\end{eqnarray} 
is a generalisation of (\ref{eq:born}) including the strong 
interaction effect. 
We have implicitly assumed above that $M_{\lambda}$ 
in (\ref{eq:mlambda}) does not contribute to the 
divergent part, because axial current conservation is broken only 
by soft operators. 

The external nucleon-line renormalization also receives 
strong interaction effects.  Using  the same notation 
${\cal M}^{(s)}$ as in (\ref{eq:mc}), while including strong interaction 
effects, 
${\cal M}^{(s)}$ is expressed in terms of 
the Green's function  (\ref{eq:green2}), 
\begin{eqnarray} 
{\cal M}^{(s)} 
&=&\frac{i}{2\sqrt{2}}G_{V}e^{2}\int \frac{d^{4}k}{ 
(2\pi )^{4}}\frac{1}{k^{2}-\lambda ^{2}} 
\bar v_{\nu}\gamma ^{\lambda }(1-\gamma ^{5})v_{e}(\ell ) 
\nonumber \\
& & 
\hskip3cm \times 
g^{\mu \nu} T_{\lambda \mu \nu}(k, p_{2}-p_{1}, p_{1}, p_{2}) 
\nonumber \\ 
& & 
+\left \{\sqrt{Z_{2}(m_{e})}-1 \right \}{\cal M}^{(0)}. 
\end{eqnarray} 
The last term is external line renormalization of the lepton. 

The integral on the r.h.s. is expressed in terms of  $T_{\mu \nu}$ 
by using (\ref{eq:c12}) and further    
$\langle p_{1}\vert t_{\mu }(0)\vert p_{2}\rangle $.  Retaining only the 
divergent part, and using (\ref{eq:highenergy1}), we find 
\begin{eqnarray} 
& &\hskip-.2cm \frac{i}{2\sqrt{2}}G_{V}e^{2}\int \frac{d^{4}k}{ 
(2\pi )^{4}}\frac{1}{k^{2}-\lambda ^{2}} 
\bar v_{\nu}\gamma ^{\lambda }(1-\gamma ^{5})v_{e}(\ell ) 
\nonumber \\ 
& &\times 
g^{\mu \nu} \left \{ 
\frac{\partial }{\partial k^{\lambda}} 
T_{\mu \nu}(-k, p_{1}, p_{2}) 
+\frac{\partial }{\partial q^{\lambda }}(M_{\mu \nu}-q^{\rho} 
\delta T_{\rho \mu \nu}) 
\Bigg \vert _{q=p_{2}-p_{1}} 
\right \} 
\nonumber \\ 
& &\sim 
\frac{i}{2\sqrt{2}}G_{V}e^{2}\int \frac{d^{4}k}{ 
(2\pi )^{4}}\frac{1}{k^{2}-\lambda ^{2}} 
\bar v_{\nu}\gamma ^{\lambda }(1-\gamma ^{5})v_{e}(\ell ) 
\nonumber \\ 
& &\times  \left \{ 
\frac{\partial }{\partial k^{\lambda}}\left ( 
\frac{2k^{\mu }}{k^{2}+2k\cdot p_{2}}\right ) 
\langle p_{1} \vert t_{\mu}(0)\vert p_{2}\rangle 
+g^{\mu \nu} 
\frac{\partial }{\partial q^{\lambda }}(M_{\mu \nu}- 
q^{\rho }\delta T_{\rho \mu \nu}) 
\Bigg \vert _{q=p_{2}-p_{1}} 
\right \}. 
\nonumber \\ 
\label{eq:firstterm} 
\end{eqnarray} 
Again, axial-current conservation is broken only by soft operators and 
the $M_{\mu \nu}$ term does not contribute to the divergent part 
of ${\cal M}^{(s)}$. Thus we can drop 
$M_{\mu \nu}-q^{\rho} \delta T_{\rho \mu \nu}$. 
We, therefore, retain only the first term 
of (\ref{eq:firstterm}) to evaluate  the divergent part of 
${\cal M}^{(s)}$: 
\begin{eqnarray} 
{\cal M}^{(s)}\Bigg \vert _{{\rm divergent}} 
&=&\frac{e^{2}}{16\pi ^{2}}\times \left (-\frac{1}{2}\right ) 
{\rm log} M^{2} {\cal M}^{(0)}+ 
\left \{ \sqrt{Z_{2}(m_{e})}-1 \right \}{\cal M}^{(0)} 
\nonumber \\ 
&\sim &-\frac{e^{2}}{16\pi ^{2}}{\rm log} M^{2} {\cal M}^{(0)}. 
\label{eq:c27} 
\end{eqnarray} 

Adding (\ref{eq:c23}) to (\ref{eq:c27}), the divergent part of 
${\cal M}^{(v)}+{\cal M}^{(s)}$ becomes 
\begin{eqnarray} 
{\cal M}^{(v)}+{\cal M}^{(s)}\Bigg \vert _{{\rm divegent}} 
=\frac{e^{2}}{16\pi ^{2}}{\rm log}M^{2} 
\left ( 
\frac{3}{2}{\cal M}^{(0)}+3\bar Q {\cal M}^{(0)} 
\right ). 
\label{eq:div} 
\end{eqnarray} 
The first part in the brackets of (\ref{eq:div})  gives divergent 
terms that are proportional to $f_{V}^{2}$ and $g_{A}^{2}$ in the 
amplitude squared, as determined by current algebra, conservation
of the vector current, and softly-broken axial current conservation.
Hence, we can conclude that the coefficient of the logarithmic divergence
$\frac{\displaystyle{3}}{\displaystyle{2}}{\rm log}(M^{2}/m_{p}^{2})$
in  (\ref{eq:in-1-F}) and that in  (\ref{eq:in-1-GT}) are both 
independent of the model of strong interactions, and smoothly
continue to the electroweak calculation. 
On the oher hand, the second term of logarithmic divergence
which arises from the interference of the vector and axial-vector
currents are model-dependent; the appearance of  $\bar Q$ is
a manifestation of the model dependence.

\section{Long-distance contributions to the 
vector-axial-vector interference terms} 
\label{sec:cpara}

To evaluate the long-distance contribution to the axial-current correction 
to the Fermi transition, Marciano and Sirlin \cite{marciano} proposed the 
prescription to make the logarithmic `divergence' milder by introducing 
nucleon form factors in the evaluation of vertex corrections. 
The result does not contain the cut-off mass $M$ any longer. 
The calculation for the Gamow-Teller transition is carried out 
parallel with that for the Fermi transition. 

Marciano and Sirlin \cite{marciano} considered only the vector 
and axial-vector current contributions, 
implicitly assuming that the contribution from 
weak magnetism is suppressed by the proton mass $O(1/m_p)$. 
We find that, since the 
loop momentum becomes of the order of the form factor mass, typically 
$m_\rho$ ($\rho$ meson mass), an $O(m_\rho/m_p)$ contribution may be 
non-negligible. 
This contribution turns out to be 
particularly large for the Gamow-Teller transition.

We work with the effective electromagnetic vertex, 
\begin{eqnarray} 
\gamma ^{\mu }F_{1}^{(p)}(k^{2})-\frac{i}{2m_{p}}\sigma ^{\mu \nu} 
k_{\nu}F_{2}^{(p)}(k^{2}), 
\end{eqnarray} 
for the proton, where $F^{(p)}_{1}(k^{2})$  and  $F^{(p)}_{2}(k^{2})$ are 
the form factors, and similarly for the neutron with the form factors, 
$F^{(n)}_{1}(k^{2})$ and $F^{(n)}_{2}(k^{2}).$ 
The form factors are also included at the weak vertex of 
the nucleons, $F_{V}((p_{2}-k-p_{1})^{2}))\approx F_{V}(k^{2})$ 
for the vector vertex and $F_{A}((p_{2}-k-p_{1})^{2}))\approx F_{A}(k^{2})$ 
for the axial-vector vertex. We also retain  in the weak vertex 
the weak magnetism term with the form factor 
$F_W(k^2)$ via 
\begin{eqnarray} 
-\frac{i}{2m_{N}}\sigma _{\lambda \rho}(p_{2}- 
k-p_{1})^{\rho}F_{W}((p_{2}-k-p_{1})^{2} 
\approx 
\frac{i}{2m_{N}}\sigma _{\lambda \rho}k^{\rho} F_{W}(k^{2}), 
\end{eqnarray} 
where $m_{N}\equiv (m_{p}+m_{n})/2$. 
The effective nucleon vertex $R_{\mu \lambda }(p_{1}, p_{2}, k)$ 
defined in (\ref{eq:rmulambda}) that contributes to ${\cal M}^{(v3)}$ 
then reads, 
\begin{eqnarray} 
R_{\mu \lambda }(p_{1}, p_{2}, k)&=& 
\frac{i}{2m_{N}}(\sigma _{\lambda \rho}k^{\rho}) 
(2p_{2}-k)_{\mu}F_{1}^{(p)}(k^{2})F_{W}(k^{2}) 
\nonumber \\ 
& & \hskip-1cm + \left \{ -i \gamma _{\lambda } 
\left ( f_{V}F_{V}(k^{2})-g_{A}\gamma ^{5}F_{A}(k^{2})\right ) + 
\frac{1}{2m_{N}}(\sigma _{\lambda \rho }k^{\rho})F_{W}(k^{2}) 
\right \} 
\nonumber \\ 
& &\hskip2cm \times (\sigma _{\mu \nu}k^{\nu})\left \{ 
F_{1}^{(p)}(k^{2})+F_{2}^{(p)}(k^{2}) 
\right \}. 
\end{eqnarray} 

We now decompose the inner part ${\cal M}^{(v3)}$ as, 
\begin{eqnarray} 
{\cal M}^{(v3)}={\cal M}_{p}^{(v3, VA)}+{\cal M}_{p}^{(v3, wm)} 
+{\cal M}_{n}^{(v3, VA)}+{\cal M}_{n}^{(v3, wm)}, 
\label{eq:1775} 
\end{eqnarray} 
where 
\begin{eqnarray} 
{\cal M}_{p}^{(v3,VA) }&=& 
\frac{1}{\sqrt{2}}G_{V}e^{2}\int \frac{d^{4}k}{(2\pi )^{4}} 
\frac{1}{(k-\ell )^{2}-m_{e}^{2}} 
\frac{1}{(p_{2}-k)^{2}-m_{p}^{2}} 
\frac{1}{k^{2}-\lambda ^{2}} 
\nonumber \\ 
& & \times \bar v_{\nu}\gamma ^{\lambda }(1-\gamma ^{5}) 
\left \{(k-2\ell )^{\mu}+i\sigma ^{\mu \nu}k_{\nu} 
\right \} v_{e}(\ell) 
\nonumber \\ 
& & \times \bar u_{n}(p_{1}) 
\gamma _{\lambda } 
\left \{ f_{V}F_{V}(k^{2})-g_{A}\gamma ^{5}F_{A}(k^{2}) \right \} 
(\sigma _{\mu \rho}k^{\rho}) 
u_{p}(p_{2}) 
\nonumber \\ 
& &\times \{F_{1}^{(p)}(k^{2})+F_{2}^{(p)}(k^{2}) \}, 
\label{eq:pv3va} 
\\ 
{\cal M}_{p}^{(v3,wm) }&=& 
\frac{i}{\sqrt{2}}G_{V}e^{2} 
\left (\frac{i}{2m_{N}}\right ) 
\int \frac{d^{4}k}{(2\pi )^{4}} 
\frac{1}{(k-\ell )^{2}-m_{e}^{2}} 
\frac{1}{(p_{2}-k)^{2}-m_{p}^{2}} 
\frac{1}{k^{2}-\lambda ^{2}} 
\nonumber \\ 
& & \times \bar v_{\nu}\gamma ^{\lambda }(1-\gamma ^{5}) 
\left \{(k-2\ell )^{\mu}+i\sigma ^{\mu \nu}k_{\nu} 
\right \} v_{e}(\ell) 
\nonumber \\ 
& & \times \bar u_{n}(p_{1})\Bigg [ 
(\sigma _{\lambda \rho}k^{\rho})(2p_{2}-k)_{\mu}F_{1}^{(p)}(k^{2}) 
\nonumber \\ 
& &-i(\sigma _{\lambda \rho}k^{\rho})(\sigma _{\mu \nu}k^{\nu}) 
\left \{ 
F_{1}^{(p)}(k^{2})+F_{2}^{(p)}(k^{2}) 
\right \} \Bigg ] 
u_{p}(p_{2}) F_{W}(k^{2})
\label{eq:pv3wm} 
\end{eqnarray} 
and likewise for the neutron, ${\cal M}_{n}^{(v3,VA)}$ and 
${\cal M}_{n}^{(v3,wm)}$. 
Eq. (\ref{eq:pv3va}) is the term due to the $V$ and $A$ current 
interaction and (\ref{eq:pv3wm}) is due to the weak magnetism. 
The calculation of the four terms in ${\cal M}^{(v3)}$ of 
(\ref{eq:1775}) is given in Appendix \ref{app:inner}.

\vskip5mm\noindent 
{\it Results for ${\cal M}_{p}^{(v3,VA)}$ and ${\cal M}_{n}^{(v3,VA)}$} 

We are interested only in  the $f_{V}g_{A}$-terms and 
we subtract the pure vector and pure axial-vector parts that are 
proportional $f_{V}^{2}$ and $g_{A}^{2}$. 
We find \begin{eqnarray} 
& &\sum _{{\rm spin}} 
\left \{ 
({\cal M}_{p}^{(v3,VA)}+{\cal M}_{n}^{(v3, VA)}){\cal M}^{(0)*} + 
({\cal M}_{p}^{(v3, VA)*}+{\cal M}_{n}^{(v3, VA)*}){\cal M}^{(0)} 
\right \} \Bigg \vert _{f_{V}g_{A}} 
\nonumber \\ 
& & \hskip0.5cm = 32 G_{V}^{2} m_{n}m_{p} E E_{\nu} 
\left (  \frac{e^{2}}{8\pi ^{2}}\right ) f_{V}g_{A}\Bigg [ 
6\left ( {\cal C}_\sigma^{(p, A)}+{\cal C}_\sigma^{(n, A)} 
\right ) (1+\beta {\rm cos}\theta ) 
\langle 1 \rangle ^{2} 
\nonumber \\ 
& & 
\hskip1cm 
+ \left ( 
6{\cal C}_\sigma^{(p, V)}+6{\cal C}_\sigma^{(n, V)}+ 
2{\cal C}_\tau^{(p, V)}+2{\cal C}_\tau^{(n, V)} \right ) 
\left (3-\beta {\rm cos}\theta \right ) 
\frac{1}{3}\langle {\mib \sigma} \rangle ^{2} 
\Bigg ], 
\label{eq:syuusei} 
\end{eqnarray} 
where ${\cal C}_{\sigma}$ and ${\cal C}_\tau$ are defined by 
the integrals 
\begin{eqnarray} 
\int \frac{d^{4}k}{(2\pi )^{4}}\frac{k_{\mu}k_{\nu}}{(k^{2})^{2} 
\{(p_{2}-k)^{2}-m_{p}^{2}\}}\left \{ 
F_{1}^{(p)}(k^{2}) + F_{2}^{(p)}(k^{2})\right \}F_{V}(k^{2}) & & 
\nonumber \\ 
\equiv 
\frac{i}{16\pi ^{2}}\left \{ 
g_{\mu \nu}{\cal C}_\sigma^{(p, V)}+\frac{1}{m_{p}^{2}}(p_{2})_{\mu } 
(p_{2})_{\nu}{\cal C}_\tau^{(p,V)} 
\right \}, & & 
\label{eq:formnum1} 
\\ 
\int \frac{d^{4}k}{(2\pi )^{4}}\frac{k_{\mu}k_{\nu}}{(k^{2})^{2} 
\{(p_{2}-k)^{2}-m_{p}^{2}\}}\left \{ 
F_{1}^{(p)}(k^{2}) + F_{2}^{(p)}(k^{2})\right \}F_{A}(k^{2}) & & 
\nonumber \\ 
\equiv 
\frac{i}{16\pi ^{2}}\left \{ 
g_{\mu \nu}{\cal C}_\sigma^{(p, A)}+\frac{1}{m_{p}^{2}}(p_{2})_{\mu } 
(p_{2})_{\nu}{\cal C}_\tau^{(p,A)} 
\right \}, & & 
\label{eq:formnum2} 
\end{eqnarray} 
and are evaluated numerically using the empirical dipole nucleon form 
factors (see Appendix \ref{app:formfactor}) as, 
\begin{eqnarray} 
& &{\cal C}_\sigma^{(p,V)}=(1+\mu _{p})\times 0.240=0.671, 
\label{eq:cspv} 
\\ 
& &{\cal C}_\tau^{(p,V)}=(1+\mu _{p})\times (-0.198)=-0.554, 
\label{eq:ctpv} 
\\ 
& &{\cal C}_\sigma^{(p,A)}=(1+\mu _{p})\times 0.261=0.729, 
\label{eq:cspa} 
\\ 
& &{\cal C}_\tau^{(p,A)}=(1+\mu _{p})\times (-0.212)=-0.592, 
\label{eq:ctpa} 
\end{eqnarray} 
where $\mu _{p}=1.793$ and $\mu _{n}=-1.913$ are the anomalous 
magnetic moments of the proton and the neutron. 

The evaluation of the neutron's counterpart 
${\cal C}_{\sigma}^{n, V}$, 
${\cal C}_{\sigma}^{n, V}$, 
${\cal C}_{\sigma}^{n, V}$ and 
${\cal C}_{\sigma}^{n, V}$ 
goes similarly, with the result, 
\begin{eqnarray} 
& &{\cal C}_\sigma^{(n,V)}=\mu _{n} \times 0.240=-0.459, 
\label{eq:csnv} 
\\ 
& &{\cal C}_\tau^{(n,V)}=\mu _{n}\times (-0.198)=0.379, 
\label{eq:ctnv} 
\\ 
& &{\cal C}_\sigma^{(n,A)}=\mu _{n}\times 0.261=-0.499, 
\label{eq:csna} 
\\ 
& &{\cal C}_\tau^{(n,A)}=\mu _{n}\times (-0.212)=0.405. 
\label{eq:ctna} 
\end{eqnarray} 

The first terms in the square brackets of (\ref{eq:syuusei}) are 
the axial-vector current 
contribution to the Fermi transition, and 
the second terms are the vector current 
contribution to the Gamow-Teller transitions. 
The computation of the former terms (i.e., those 
proportional to 
$\langle 1\rangle^2$) 
was already done by the authors of \cite{marciano} 
and  \cite{towner}, and our evaluation 
\begin{eqnarray} 
f_{V}g_{A}\times 6\left ( {\cal C}_\sigma^{(p, A)}+{\cal C}_\sigma^{(n, A)} 
\right )=2\times 0.875=1.75 
\label{eq:m-s-t} 
\end{eqnarray} 
agrees with their results allowing for the difference 
that arises from the different values of the input data 
\footnote{ 
Our value 0.875 in (\ref{eq:m-s-t}) 
should be compared with 0.885 of Marciano and Sirlin 
\cite{marciano} and with 0.881 of Towner \cite{towner}. 
The slight difference is due to different choices of $g_{A}$ and 
$m_{A}$ in the form factor. 
}. 

Let us remark that 
(\ref{eq:cspv}) - (\ref{eq:ctpa}) and (\ref{eq:csnv}) - 
(\ref{eq:ctna}) reduce to 
\begin{eqnarray} 
& &{\cal C}_\sigma^{(p,V)}, {\cal C}_\sigma^{(p, A)}\longrightarrow 
\frac{1}{4}{\rm log}\left (\frac{M^{2}}{m_{p}^{2}}\right )+\frac{3}{8}, 
\\ 
& &{\cal C}_\sigma^{(n, V)}, {\cal C}_\sigma^{(n, A)} \longrightarrow 0, 
\\ 
& &{\cal C}_\tau^{(p, V)}, {\cal C}_\tau^{(p, A)} 
\longrightarrow -\frac{1}{2}, 
\\ 
& &{\cal C}_\tau^{(n, V)}, {\cal C}_\tau^{(n, A)} \longrightarrow 0, 
\end{eqnarray} 
for the point-like nucleon with a vanishing anomalous magnetic moment, and 
(\ref{eq:syuusei}) reduces 
to the $f_{V}g_{A}$-terms in (\ref{eq:a3}).

\vskip5mm\noindent 
{\it Results for ${\cal M}_{p}^{(v3, wm)}$ and ${\cal M}_{n}^{(v3, wm)}$} 

The calculation is also straightforward for 
the effects of weak magnetism, though it is rather lengthy. 
We refer the readers who are interested 
in the calculation to Appendix \ref{app:inner}; 
here we only record the final result: 
\begin{eqnarray} 
& &\sum _{{\rm spin}} \left \{ \left ({\cal M}_{p}^{(v3, wm)} 
+{\cal M}_{n}^{(v3, wm)}\right ){\cal M}^{(0)*} 
+ 
\left ( {\cal M}_{p}^{(v3, wm)*} 
+{\cal M}_{n}^{(v3, wm)*}\right ) {\cal M}^{(0)}\right \} 
\nonumber \\ 
& =& 
32G_{V}^{2}m_{n}m_{p}EE_{\nu}\left ( \frac{e^{2}}{8\pi ^{2}}\right ) 
\frac{1}{m_{N}}\Bigg [ 
2g_{A} 
\left (m_{p}{\cal D}_\sigma^{(p)}+m_{n}{\cal D}_\sigma^{(n)}\right ) 
\left (3-\beta {\rm cos}\theta \right ) 
\frac{1}{3}\langle {\mib \sigma}  \rangle ^{2} 
\nonumber \\ 
& & 
+g_{A}\left (m_{p}{\cal E}^{(p)} - m_{n}{\cal E}^{(n)}\right ) 
(3-\beta {\rm cos}\theta )\frac{1}{3}\langle {\mib \sigma } \rangle 
^{2}
\nonumber \\ 
& & + 
\frac{3}{2}f_{V}\left ( m_{p}{\cal E}^{(p)}+m_{n}{\cal E}^{(n)}\right ) 
(1+\beta {\rm cos}\theta )\langle 1 \rangle ^{2} 
\Bigg ],
\label{eq:2268} 
\end{eqnarray} 
where the constants ${\cal D}_{\sigma }^{(p)}$  and ${\cal E}^{(p)}$ 
are defined by 
\begin{eqnarray} 
& &\int \frac{d^{4}k}{(2\pi )^{4}} 
\frac{k^{\mu }k^{\nu}}{(k^{2})^{2}}\frac{1}{(p_{2}-k)^{2}-m_{p}^{2}} 
F_{W}(k^{2})F_{1}^{(p)}(k^{2}) 
\nonumber \\
& & \hskip4cm \equiv 
\frac{i}{16\pi ^{2}}\left \{ 
g^{\mu \nu}{\cal D}^{(p)}_\sigma+\frac{1}{m_{p}^{2}}p_{2}^{\mu } 
p_{2}^{\nu}{\cal D}_\tau^{(p)}\right \}, 
\label{eq:dspdtp} 
\\ 
& &\int \frac{d^{4}k}{(2\pi )^{4}} 
\frac{k^{\mu}}{k^{2}}\frac{1}{(p_{2}-k)^{2}-m_{p}^{2}} 
F_{W}(k^{2})\left \{F_{1}^{(p)}(k^{2})+F_{2}^{(p)}(k^{2})\right \}
\nonumber \\
& & \hskip4cm \equiv 
\frac{i}{16\pi ^{2}}p_{2}^{\mu}{\cal E}^{(p)}, 
\label{eq:e(p)} 
\end{eqnarray} 
and are evaluated, by using the phenomenological form factors 
of the proton in Appendix \ref{app:formfactor}, as 
\begin{eqnarray} 
{\cal D}_\sigma^{(p)}&=&(\mu _{p}-\mu _{n})\times 0.240 
=0.890, 
\\ 
{\cal D}_\tau^{(p)}&=&(\mu _{p}-\mu _{n})\times (-0.198) 
=-0.735, 
\\ 
{\cal E}^{(p)}&=&(\mu _{p}-\mu _{n})(1+\mu _{p})\times 0.0836 
=0.866. 
\end{eqnarray} 
[Note that ${\cal D}_\tau^{(p)}$ does not appear in (\ref{eq:2268}).] 
We also define the corresponding quantities 
for the neutron and obtain an expression similar to (\ref{eq:2268}), where 
\begin{eqnarray} 
{\cal D}_\sigma^{(n)}&=&0, 
\\ 
{\cal D}_\tau^{(n)}&=&0, 
\\ 
{\cal E}^{(n)}&=&(\mu _{p}-\mu _{n})\mu _{n} \times 0.0835 
=-0.592. 
\end{eqnarray} 
\vskip5mm 
\noindent 
{\it Summary of long-distance contributions} 

The long-distance contributions of ${\cal M}^{(v3)}$ 
defined by (\ref{eq:1775}) is summarized as 
\begin{eqnarray} 
& &\hskip-2cm\sum _{{\rm spin}} \left ( 
{\cal M}^{(v3)}{\cal M}^{(0)*}+{\cal M}^{(v3)*}{\cal M}^{(0)} 
\right ) 
\nonumber \\ 
&=& 
32G_{V}^{2}m_{n}m_{p}EE_{\nu}\cdot \frac{e^{2}}{8\pi ^{2}} 
\Bigg [ 
C^{\rm F}(1+\beta\cos\theta) 
f_{V}^{2}\langle 1 \rangle ^{2} 
\nonumber \\ 
& & 
+C^{\rm GT}(3-\beta {\rm cos}\theta) 
\frac{1}{3}g_{A}^{2}\langle {\mib \sigma } \rangle ^{2} 
\Bigg ]. 
\end{eqnarray} 
Here the constants $C_{}^{\rm F}$ and $C_{}^{\rm GT}$, which were introduced 
in (\ref{eq:summe}) and (\ref{eq:d-in-gt}), 
are obtained from (\ref{eq:syuusei}) and (\ref{eq:2268}): 
\begin{eqnarray} 
f_{V}^{2}C_{}^{\rm F}&=&6f_{V}g_{A}\left ( 
{\cal C}_\sigma^{(p, A)}+{\cal C}_\sigma^{(n, A)} 
\right )+ \frac{3 f_{V}}{2m_{N}}\left ( 
m_{p}{\cal E}^{(p)}+m_{n}{\cal E}^{(n)}\right ), 
\\ 
g_{A}^{2} C_{}^{\rm GT}&=&f_{V}g_{A}\left ( 
6{\cal C}_\sigma^{(p, V)}+ 
6{\cal C}_\sigma^{(n, V)}+2{\cal C}_\tau^{(p, V)}+2{\cal C}_\tau^{(n, V)} 
\right ) 
\nonumber \\ 
& & \hskip-0.5cm +\frac{g_{A}}{m_{N}}\left \{2 \left ( 
m_{p}{\cal D}_\sigma^{(p)}+m_{n}{\cal D}_\sigma^{(n)} 
\right )+\left ( 
m_{p}{\cal E}^{(p)}-m_{n}{\cal E}^{(n)} 
\right )\right \}, 
\end{eqnarray} 
which are evaluated numerically as, 
\begin{eqnarray} 
C_{}^{\rm F}=1.751+0.409=2.160, 
\cr
C_{}^{\rm GT}=0.727+2.554= 3.281, 
\label{eq:coeff} 
\end{eqnarray} 
where the two parts represent contributions from 
the (V,A) interaction and weak magnetism. 
Note that the weak magnetism contribution is non-negligible for
$C_{}^{\rm F}$, and is even larger than the $(V,A)$ contributions in
$C_{}^{\rm GT}$.

\section{Summary}

Full one-loop radiative corrections are calculated for neutrino-nucleon 
quasi-elastic scattering for both Fermi and Gamow-Teller 
transitions. We separate the corrections into the outer 
and inner parts {\` a} la Sirlin. The outer part is infrared and ultraviolet 
finite, and involves the positron  velocity. This part takes different 
forms for angular independent and dependent parts of the differential 
cross section. The calculation of the inner part requires
a scrutiny regarding the continuation of the long-distance hadronic 
calculation to the short-distance quark treatment and the dependence on the 
model of hadron structure. We have shown 
that the logarithmically divergent parts do not depend on the 
structure of hadrons  not only for the Fermi part, 
but also for the Gamow-Teller 
part. This observation has enabled us to deal with the inner part for the 
Gamow-Teller transition in a way parallel to that for the Fermi 
transition. The inner corrections contribute to the 
angular-independent and dependent parts  of the cross section in the 
same manner, so that they are written as universal 
multiplicative factors on the 
coupling constants, $f_V$ and $g_A$ \cite{sirlin}.

The resulting radiative correction to the differential cross section 
(\ref{eq:diffcross}) takes the 
factorised form: 
\begin{eqnarray} 
A(\beta) &=& \left \{ 1+\delta_{\rm out}(E)\right \} 
      \left(\bar f_V^2 \langle 1 \rangle^2 
     +\bar g_A^2 \langle {\mib \sigma} \rangle^2\right)\ , 
\label{eq:summaryA} 
\\ 
B(\beta) &=& \left \{ 1+\tilde\delta_{\rm out}(E)\right \} 
      \left(\bar f_V^2\langle 1 \rangle^2 
     -\frac{1}{3}\bar g_A^2\langle {\mib \sigma} \rangle^2\right)\ , 
\label{eq:summaryB} 
\end{eqnarray} 
where $\delta_{\rm out}(E)$ and $\tilde\delta_{\rm out}(E)$ are given by 
(\ref{eq:d-out}) and (\ref{eq:d-out2}), respectively, and the inner 
corrections are absorbed into the modification of the coupling constants, 
\begin{eqnarray} 
\bar f_V^2 &=& f_V^2 \left (1+\delta_{\rm in}^{\rm F}\right ),  \cr 
\bar g_A^2 &=& g_A^2 \left( 1+\delta_{\rm in}^{\rm GT}\right )\ , 
\end{eqnarray} 
with  $\delta_{\rm in}^{\rm F}$ and 
$\delta_{\rm in}^{\rm GT}$ given  by 
\begin{eqnarray} 
\delta_{\rm in}^{\rm F} &=& \frac{\alpha}{2\pi}\left \{4\log 
\left ( \frac{m_Z}{m_p}\right ) 
+ \log \left (\frac{m_p}{M}\right ) + C^{\rm F}\right \} ,  \\ 
\delta_{\rm in}^{\rm GT}&=& \frac{\alpha}{2\pi}\left \{ 
4\log \left (\frac{m_Z}{m_p}\right ) 
+\log \left (\frac{m_p}{M}\right ) + 1 + C^{\rm GT}\right \} , 
\end{eqnarray} 
and the constants $C^{\rm F,GT}$ by (\ref{eq:coeff}) [$\alpha$ is the 
fine structure constant, $\alpha =e^{2}/(4\pi )$]. 
If we set the mass scale, $M$, at which 
the short-distance asymptotic behaviour onsets, to be 1 GeV, 
we obtain 
\begin{eqnarray} 
\delta_{\rm in}^{\rm F} &=& 
0.02370 - 1.16\times 10^{-3} \times {\rm log}\left ( 
\frac{M}{1{\rm GeV}} 
\right ) 
\nonumber \\ 
&=& 0.02370 \pm0.0008, \cr 
\delta_{\rm in}^{\rm GT} &=& 
0.02616-1.16\times 10^{-3} \times {\rm log}\left ( 
\frac{M}{1{\rm GeV}} 
\right ) 
\nonumber \\ 
&=& 
0.02616 \pm 0.0008, 
\end{eqnarray} 
the error corresponding to a multiplicative factor of 2 in the scale 
$M$. 

\begin{figure}[htbp] 
\begin{center} 
\includegraphics*[scale=.55]{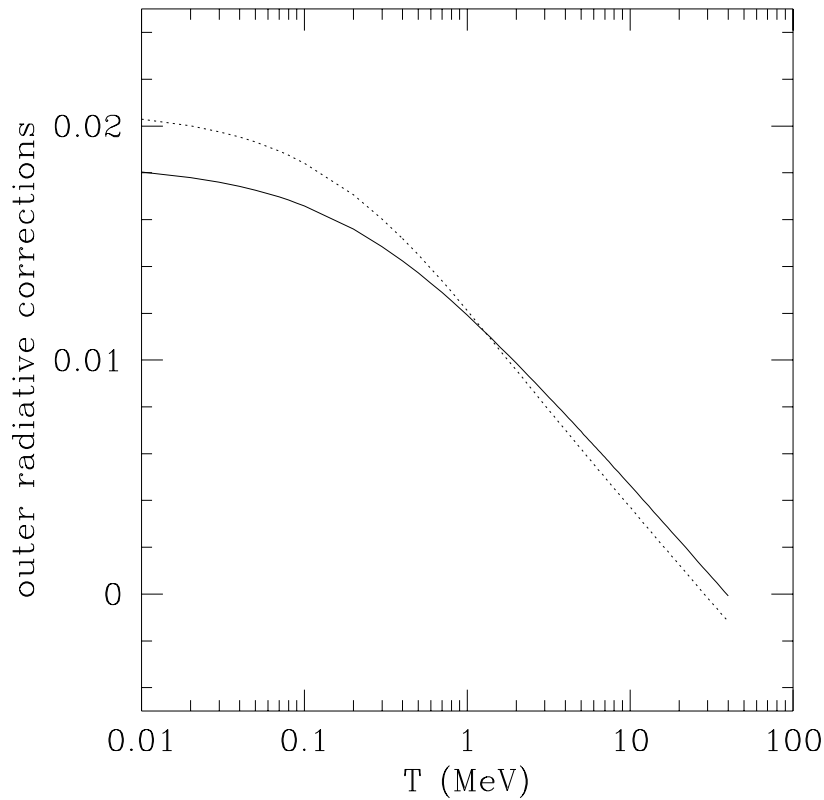} 
\end{center} 
\caption{
The outer corrections, 
$\delta _{\rm out} (E)$ (solid line) and 
$\tilde \delta _{\rm out}(E)$ (dotted line) 
as a function of the kinetic energy   
$T=E-m_{e}$ of the recoil positron.} 
\label{fig:outer} 
\end{figure}

We present numerical values of $\delta_{\rm out}(E)$ 
and $\tilde \delta _{\rm out} (E)$ in Table 1 
and Figure \ref{fig:outer}, where $T=E-m_{e}$ is the 
kinetic energy of the positron. 
In particular, at the threshold $E=m_{e}$, they are given  
by 
\begin{eqnarray} 
\delta _{\rm out}(m_{e})&=&\frac{e^{2}}{8\pi ^{2}} 
\left \{ 
-\frac{27}{4}+\frac{3}{2}{\rm log}\left ( 
\frac{m_{p}^{2}}{m_{e}^{2}}\right )\right \} 
=0.01835, 
\\ 
\tilde \delta _{\rm out}(m_{e})&=&\frac{e^{2}}{8\pi ^{2}} 
\left \{ 
-\frac{19}{4}+\frac{3}{2}{\rm log}\left ( 
\frac{m_{p}^{2}}{m_{e}^{2}}\right )\right \} 
=0.02067, 
\end{eqnarray} 
and for a large positron energy, they behave as 
\begin{eqnarray} 
\delta _{\rm out}(E) \sim \frac{e^{2}}{8\pi ^{2}} 
\left \{ 
\frac{23}{4}-\frac{4\pi ^{2}}{3}+3{\rm log}\left ( 
\frac{m_{p}}{2E}\right )\right \}, 
\\ 
\tilde \delta _{\rm out}(E) \sim \frac{e^{2}}{8\pi ^{2}} 
\left \{ 
\frac{19}{4}-\frac{4\pi ^{2}}{3}+3{\rm log}\left ( 
\frac{m_{p}}{2E}\right )\right \}. 
\end{eqnarray} 
$\delta _{\rm out}(E)$ and $\tilde \delta _{\rm out}(E)$ 
differ only by $\approx 10$\%, and $\delta_{\rm in}^{\rm F}$ 
and $\delta_{\rm in}^{\rm GT}$ 
also differ by 10\%; hence the radiative correction to the 
angular dependence is at a 0.1\% level.

\vskip0.5cm 

\begin{center} 
\begin{tabular}{ r  c  r  r } 
\hline\hline 
$T$ (MeV) & $\beta $ & $\delta _{\rm out}(E)$ & 
$\tilde \delta _{\rm out}(E)$ 
\\ 
\hline 
0.01 & 0.1950 & 0.0180 & 0.0203 
\\ 
0.05 & 0.4127 & 0.0173 & 0.0193 
\\ 
0.1 & 0.5482 & 0.0166 & 0.0184 
\\ 
0.2 & 0.6953 & 0.0156 & 0.0171 
\\ 
0.3 & 0.7765 & 0.0149 & 0.0160
\\ 
0.4 & 0.8279  & 0.0142 & 0.0152 
\\ 
0.5 & 0.8629  & 0.0137 & 0.0145 
\\ 
0.6 & 0.8879 & 0.0132 & 0.0139 
\\ 
0.7 & 0.9066 & 0.0129 & 0.0134 
\\ 
0.8 & 0.9209 & 0.0125 & 0.0129 
\\ 
0.9 & 0.9321 & 0.0122 & 0.0125 
\\ 
1.0 & 0.9411 & 0.0119 & 0.0121 
\\ 
2.0 & 0.9791 & 0.0099 & 0.0096 
\\ 
3.0  & 0.9894 & 0.0086 & 0.0081 
\\ 
4.0  & 0.9936 & 0.0077 & 0.0070 
\\ 
5.0 & 0.9957 & 0.0070 & 0.0062 
\\ 
6.0 & 0.9969 & 0.0064 & 0.0055 
\\ 
7.0 &  0.9977 & 0.0058 & 0.0500 
\\ 
8.0 & 0.9982 & 0.0054 & 0.0045 
\\ 
9.0 & 0.9986 & 0.0050 & 0.0041 
\\ 
10.0 & 0.9988 & 0.0047 & 0.0037 
\\ 
20.0 &  0.9997 & 0.0023 & 0.0013 
\\ 
30.0 & 0.9999 & 0.0009 & $-$0.0002 
\\ 
40.0 & 0.9999 & $-$0.0001 & $-$0.0012 
\\ 
\hline 
\end{tabular} \end{center} 

\vskip0.2cm 
\noindent 
Table 1: Numerical values of the outer corrections, 
$\delta _{\rm out} (E)$ and $\tilde \delta _{\rm out}(E)$ 
as a function of the kinetic energy  $T=E-m_{e}$ of the 
recoil positron. 
\vskip0.5cm 

\vskip0.5cm 
The dominant part of the inner correction 
arises from $4\log (m_Z/m_p)$, 
which amounts to 18.18. The change of the 
scale 
$M=1$ GeV in $\log (m_p/M)$ by a factor of 2 leads to an error of $\pm 0.69$. 
The model-dependent, 
long-distance contribution from axial-vector-vector interference terms 
contributes by 2.160 for the Fermi transition, and 3.281 for the 
Gamow-Teller 
transition (there is an additional contribution of unity for the  Gamow-Teller 
transition). We find a sizable contribution to the 
axial-vector-vector interference 
terms from the weak magnetism, which has been ignored in \cite{marciano} and
\cite{towner}. 
This contribution is even larger than that 
of the ($V,A$) interaction for the correction to the Gamow-Teller 
transition. 
If we would use the point nucleon calculation for the 
Gamow-Teller transition as an extreme case 
the long-distance contribution to $C^{\rm GT}$ will be 1.1. 
We may, therefore, conclude that the uncertainty from the model 
dependence is conservatively no more than 10\% for $\delta_{\rm in}$.

The formulae we derived are applicable not only to 
neutrino-nucleon 
quasi elastic scattering, but also to other neutrino reactions off nuclei. 

The radiative correction to nuclear beta decay is also described by 
(\ref{eq:summaryA}) with $\delta_{\rm out}$ replaced by the well-known 
$g$ function of \cite{kinoshitasirlin,sirlin} 
\begin{eqnarray} 
\hat \delta_{\rm out}&=&\frac{\alpha}{2\pi}g(E,E_0)
\nonumber \\ 
&=& \frac{\alpha}{2\pi}\Bigg[ 
3\  {\rm ln} \left( {m_p \over m_e}\right) 
- {3 \over 4} + {4 \over \beta } L 
\left({{2\beta }\over {1+\beta }}\right) \nonumber \\ 
& &  + 4 \left (\frac{1}{\beta }{\rm tanh}^{-1} \beta  -1 \right ) 
\left[{{E_0 -E} \over {3E}} - {3 \over 2} + {\rm ln} 
{{2(E_0 -E) } \over m_e}\right]\nonumber \\ 
& & + { 1 \over \beta } {\rm tanh}^{-1} \beta 
\left \{2(1+\beta ^2) 
+ {{(E_0 -E)^2} \over {6E^2}} - 4 {\rm tanh}^{-1} \beta  \right \}\Bigg]~~, 
\label{eq:3.6.18} 
\end{eqnarray} 
where $E_0$ is the end point energy of the electron. 
The difference from the case of $0^+\rightarrow 0^+$ decay is that we have 
now somewhat larger inner part correction for the coupling constant that 
is multiplied on the Gamow-Teller matrix element. For example, the 
neutron decay width is given by 
\begin{equation} 
\Gamma=\frac{G_V^2}{2\pi^3}m_e^5f(E_0)\left[\bar f_V^2\langle 1\rangle^2+ 
\bar g_A^2\langle \sigma\rangle^2\right], 
\end{equation} 
where $f(E_0)=1.71483$ \cite{wilkinson} 
includes the outer radiative correction. The value in the square bracket 
is 5.966, which is 0.20\% larger than  5.954 that would be obtained by 
assuming 
$\delta_{\rm in}^{\rm GT}$ would equal $\delta_{\rm in}^{\rm F}$ 
\cite{marciano2}, where $g_A=1.2670$ \cite{pdg} is used as a 
provisional value. 
Unfortunately, the values of $g_A$ reported in the literature or
in the table of Particle Data Group \cite{pdg} are neither
the bare value nor the value including the outer radiative 
corrections defined through the prescription discussed in this
paper. Therefore, the current accuracy of $g_A$ does not warrant 
the use of our formulae, e.g., to determine precisely
$|U_{ud}|=\cos\theta_C$\footnote{Particle Data 
Group \cite{pdg} gives a value $g_A=1.2670\pm0.0030$. This value is 
obtained by naively averaging over values reported in the literature, 
one of which is obtained including the outer radiative correction 
(up to an unkown constant; the inner correction is ignored), 
and some others are results which 
do not include radiative corrections at all. Unfortunately, the 
value with radiative correction reported by \cite{abele} differs 
 from others that ignore radiative corrections \cite{glueck} by
an amount much larger 
than the difference that would be accounted for with radiative 
corrections, 
and hence much of
this difference is probably ascribed to other systematic errors 
of experiments. For this reason we give up to apply corrections to find 
a bare $g_A$ from the available data.}, and 
we should postpone a detailed numerical analysis until an accurate 
estimate of $g_A$ will become available. 

Our final remark is that one may not necessarily need the formulae 
we derived for the purpose to obtain a precise estimate of neutrino 
nucleon scattering cross section or of $|U_{ud}|$, if $g_A$ would
be properly obtained by using the outer-part radiative correction formula
in which the inner and outer corrections are accurately separated 
and the inner corrections are included in $g_A$. 
The true power of our formulae 
emerges when one would deal radiative corrections for
processes involving the neutral current.


\vskip1cm 
\noindent 
{\large {\bf Acknowledgement}} 

We would like to thank Professor A. Sirlin for his useful communication. 
The present work is supported in part by Grants in Aid 
of the Ministry of Education.

\vskip20mm 


\appendix 
\section{Calculation of ${\cal M}^{(v2)}$ in (\ref{eq:ma2})} 
\label{app:mv2}

We sketch the calculation of ${\cal M}^{(v2)}$ in (\ref{eq:ma2}). 
The integration formula necessary for 
(\ref{eq:ma2}) is given in Appendix of Ref.  \cite{abers}, \begin{eqnarray} 
& & \hskip-2cm 
i\int \frac{d^{4}k}{(2\pi )^{4}}\frac{1}{(k-\ell )^{2}-m_{e}^{2}} 
\frac{1}{(k-p_{2})^{2}-m_{p}^{2}}\frac{1}{k^{2}-\lambda ^{2}}k_{\mu} 
\nonumber \\ 
&=&\frac{1}{16\pi ^{2}}\frac{1}{m_{p}^{2}}\Bigg [ 
\left \{ 
-\frac{1}{\beta }{\rm tanh}^{-1}\beta +\frac{1}{2}{\rm log}\left ( 
\frac{m_{p}^{2}}{m_{e}^{2}}\right ) 
\right \}(p_{2})_{\mu} 
\nonumber \\ 
& & \hskip1.5cm + \left \{ 
\frac{1}{E\beta }(m_{p}+E){\rm tanh}^{-1}\beta -\frac{1}{2}{\rm log} 
\left ( 
\frac{m_{p}^{2}}{m_{e}^{2}} 
\right ) 
\right \}\ell _{\mu} 
\Bigg ]. 
\label{eq:jal1} 
\end{eqnarray} 
Putting  (\ref{eq:jal1}) into (\ref{eq:ma2}) and taking the nucleon static 
limit, we get 
\begin{eqnarray} 
{\cal M}^{(v2)}&=&\frac{G_{V}}{\sqrt{2}}\left ( \frac{e^{2}}{8\pi ^{2}} 
\right ) \frac{1}{E \beta }{\rm tanh}^{-1}\beta \:\:\: 
\nonumber \\ 
& &\times  [ \bar v_{\nu}\gamma ^{\lambda }(1-\gamma ^{5}) 
i\sigma ^{0 \nu}v_{e}(\ell ) ] \ell _{\nu} 
[ \bar u_{n}(p_{1})W_{\lambda }(p_{1}, p_{2})u_{p}(p_{2}) ]\ , 
\end{eqnarray} 
where only the term proportional to  $m_{p}$ in (\ref{eq:jal1}) 
is retained. The spin summation of the tree-one-loop interference 
amplitude is expressed as 
\begin{eqnarray} 
& & \hskip-1cm \sum _{\rm spin}\left ( 
{\cal M}^{(v2)}{\cal M}^{(0)*}+{\cal M}^{(v2)*}{\cal M}^{(0)} 
\right )
\nonumber \\
& & =\left ( \frac{G_{V}}{\sqrt{2}} \right )^{2} 
\left ( \frac{e^{2}}{4\pi ^{2}} \right ) \frac{1}{E\beta } 
{\rm tanh}^{-1}\beta \times Q^{\lambda \rho}K_{\lambda \rho}, 
\end{eqnarray} 
where 
\begin{eqnarray} 
K_{\lambda \rho}&=&\sum _{\rm spin}[\bar u_{n}(p_{1})\gamma _{\lambda } 
(f_{V}-g_{A}\gamma ^{5})u_{p}(p_{2})] 
[\bar u_{n}(p_{1})\gamma _{\rho } 
(f_{V}-g_{A}\gamma ^{5})u_{p}(p_{2})]^{*}, 
\nonumber \\ 
Q^{\lambda \rho}&=&\sum _{\rm spin} 
[ \bar v_{\nu}\gamma ^{\lambda }(1-\gamma ^{5}) 
i\sigma ^{0 \nu}v_{e}(\ell ) ] \ell _{\nu} [ 
\bar v_{\nu}\gamma ^{\rho}(1-\gamma ^{5})v_{e}(\ell )]^{*}, 
\end{eqnarray} 
are evaluated as 
\begin{eqnarray} 
& &K_{00}=8f_{V}^{2}m_{n}m_{p}, \hskip0.5cm 
K_{ij}=8g_{A}^{2}m_{n}m_{p}\delta _{ij}, 
\\ 
& &Q^{00}=8E^{2}E_{\nu}\beta(\beta+{\rm cos}\theta ), 
\hskip0.5cm 
Q^{ij}\delta _{ij}=8E^{2}E_{\nu}\beta (3\beta-{\rm cos}\theta ). 
\end{eqnarray} 
We arrive at 
\begin{eqnarray} 
& & \sum _{\rm spin}\left ( 
{\cal M}^{(v2)}{\cal M}^{(0)*}+{\cal M}^{(v2)*}{\cal M}^{(0)} 
\right ) 
\nonumber \\ 
&=&32G_{V}^{2}m_{n}m_{p}EE_{\nu}\left ( 
\frac{e^{2}}{4\pi ^{2}}\right ){\rm tanh}^{-1}\beta 
\left \{ 
f_{V}^{2}(\beta+{\rm cos}\theta )+g_{A}^{2}(3\beta- 
{\rm cos}\theta ) \right \}, 
\nonumber \\
\end{eqnarray} 
which is the ${\cal M}^{(v2)}$ part of (\ref{eq:311}).

\section{Calculation of ${\cal M}^{(v3)}$ in (\ref{eq:ma3})} 
\label{app:mv3}

Neglecting $\lambda $ and $m_{e}$, ${\cal M}^{(v3)}$ is written as 
\begin{eqnarray} 
{\cal M}^{(v3)}&=&\frac{G_{V}}{\sqrt{2}}e^{2}\int \frac{d^{4}k}{( 
2\pi )^{4}}\frac{1}{(k^{2})^{2}}\frac{1}{(p_{2}-k)^{2}-m_{p}^{2}} 
\bar v _{\nu} \gamma ^{\lambda }(1-\gamma ^{5})i\sigma ^{\mu \nu} 
k_{\nu}v_{e}(\ell ) 
\nonumber \\ 
& & \times \bar u_{n}(p_{1})\gamma _{\lambda }(f_{V}-g_{A}\gamma ^{5}) 
\sigma _{\mu \rho}k^{\rho}u_{p}(p_{2}) 
\nonumber \\ 
&=& 
-\frac{G_{V}}{\sqrt{2}}\left (\frac{e^{2}}{16\pi ^{2}}\right ) 
{I_{\nu}}^{\rho}{X^{\nu}}_{\rho}\ , 
\end{eqnarray} 
where 
\begin{eqnarray} 
\frac{i}{16\pi ^{2}} {I_{\nu}}^{\rho}
& \equiv & 
\int \frac{d^{4}k}{(2\pi )^{4}}\frac{1}{(k^{2})^{2}} 
\frac{1}{(p_{2}-k)^{2}-m_{p}^{2}}k_{\nu}k^{\rho} 
\nonumber \\ 
&=& 
\frac{i}{16\pi ^{2}} \left \{ 
g_{\nu }^{\rho}\left [\frac{1}{4}{\rm log}\left (\frac{ 
M^{2}}{m_{p}^{2}}\right )+\frac{3}{8}\right ]+ 
\frac{1}{m_{p}^{2}}(p_{2})_{\nu}(p_{2})^{\rho}\times 
\left (-\frac{1}{2}\right )\right \}\ , 
\nonumber \\
\end{eqnarray} 
and 
\begin{eqnarray} 
{X^{\nu}}_{\rho}= 
[\bar v_{\nu}\gamma ^{\lambda }(1-\gamma ^{5})\sigma ^{\mu \nu} 
v_{e}(\ell )] 
[\bar u_{n}(p_{1}) \gamma _{\lambda }(f_{V}-g_{A}\gamma ^{5}) 
\sigma _{\mu \rho}u_{p}(p_{2})]. 
\label{eq:xnurho} 
\end{eqnarray} 
We shuffle $\gamma_\lambda\sigma_{\mu \rho}$ in 
(\ref{eq:xnurho}) using \begin{eqnarray} 
\gamma ^{\lambda} \gamma ^{\mu } \gamma ^{\nu}=g^{\lambda \mu} 
\gamma ^{\nu}+ g^{\mu \nu}\gamma ^{\lambda }-g^{\lambda \nu}\gamma ^{\mu} 
+i\varepsilon ^{\lambda \mu \nu \rho}\gamma _{\rho}\gamma ^{5} 
\end{eqnarray} 
in such a way that a single gamma matrix is sandwiched between 
spinors; we find in the nucleon static limit 
\begin{eqnarray} 
g^{\nu \rho}X_{\nu \rho} 
&=& 
-6(f_{V}+g_{A})[\bar v_{\nu}\gamma ^{0}(1-\gamma ^{5}) 
v_{e}(\ell )][\bar u_{n}(p_{1})\gamma _{0}u_{p}(p_{2})] 
\nonumber \\ 
& & +6(f_{V}+g_{A})[\bar v_{\nu}\gamma ^{i}(1-\gamma ^{5})v_{e}(\ell )][ 
\bar u_{n}(p_{1})\gamma _{i}\gamma ^{5}u_{p}(p_{2})], 
\nonumber \\
\label{eq:gnurhoxnurho} 
\\ 
\frac{1}{m_{p}^{2}}(p_{2})^{\nu}(p_{2})^{\rho}X_{\nu \rho} 
&=& 
-3f_{V}[\bar v _{\nu}\gamma ^{0}(1-\gamma ^{5})v_{e}(\ell )][ 
\bar u_{n}(p_{1})\gamma _{0}u_{p}(p_{2})] 
\nonumber \\ 
& &+ 
(2f_{V}+g_{A})[\bar v_{\nu}\gamma ^{i}(1-\gamma ^{5})v_{e}(\ell )][ 
\bar u_{n}(p_{1})\gamma _{i}\gamma ^{5}u_{p}(p_{2})]. 
\nonumber \\
\label{eq:ppxnurho} 
\end{eqnarray} 
The first and second terms  correspond to the Fermi and Gamow-Teller 
parts, respectively, in (\ref{eq:gnurhoxnurho}) 
and (\ref{eq:ppxnurho}). The angular distribution is 
determined  by the leptonic parts, 
\begin{eqnarray} 
\sum _{\rm spin} 
[\bar v_{\nu}\gamma ^{0}(1-\gamma ^{5})v_{e}(\ell )][ 
\bar v_{\nu}\gamma _{0}(1-\gamma ^{5})v_{e}(\ell )] 
&=&8EE_{\nu}(1+\beta {\rm cos}\beta ), 
\\ 
\sum _{\rm spin} 
[\bar v_{\nu}\gamma ^{i}(1-\gamma ^{5})v_{e}(\ell )][ 
\bar v_{\nu}\gamma ^{j}(1-\gamma ^{5})v_{e}(\ell )]\delta _{ij} 
&=&8EE_{\nu}(3-\beta {\rm cos}\beta ). 
\end{eqnarray} 
Combining with the nucleon parts, \begin{eqnarray} 
\sum _{\rm spin}[\bar u_{n}(p_{1})\gamma ^{0}u_{p}(p_{2})] 
[\bar u_{n}(p_{1})\gamma _{0}u_{p}(p_{2})]^{*}=8m_{n}m_{p}, \\ 
\sum _{\rm spin}[\bar u_{n}(p_{1})\gamma _{i} \gamma ^{5}u_{p}(p_{2})] 
[\bar u_{n}(p_{1})\gamma _{j}\gamma ^{5}u_{p}(p_{2})]^{*}= 
-8m_{n}m_{p}g _{ij}, 
\end{eqnarray} 
we end up with the ${\cal M}^{(v3)}$ contribution to the 
differential cross section given by, 
\begin{eqnarray} 
& & \hskip-1cm \sum _{\rm spin}\left ( 
{\cal M}^{(v3)}{\cal M}^{(0)*}+{\cal M}^{(v3)*}{\cal M}^{(0)} 
\right ) 
\nonumber \\ 
& & =32G_{V}^{2}m_{n}m_{p}EE_{\nu}\left ( 
\frac{e^{2}}{8\pi ^{2}}\right ) 
(1+\beta {\rm cos}\beta ) 
\nonumber \\ 
& & \hskip0.5cm \times \left [6f_{V}\left (f_{V}+g_{A}\right ) \left \{ 
\frac{1}{4}{\rm log}\left ( 
\frac{M^{2}}{m_{p}^{2}}\right )+\frac{3}{8}\right \} 
+3f_{V}^{2}\cdot 
\left ( -\frac{1}{2}\right ) \right ] 
\nonumber \\ 
& & +32G_{V}^{2}m_{n}m_{p}EE_{\nu}\left ( 
\frac{e^{2}}{8\pi ^{2}}\right ) 
(3-\beta {\rm cos}\beta ) 
\nonumber \\ 
& & \hskip0.5cm 
\times \left [6g_{A}\left (f_{V}+g_{A}\right ) \left \{ 
\frac{1}{4}{\rm log}\left ( 
\frac{M^{2}}{m_{p}^{2}}\right )+\frac{3}{8}\right \} 
+g_{A}(2f_{V}+g_{A})\cdot 
\left ( -\frac{1}{2}\right ) \right ]. 
\nonumber \\
\label{eq:B11} 
\end{eqnarray} 
This is (\ref{eq:a3}).  The first and second 
terms correspond to the Fermi and Gamow-Teller contributions, 
respectively [the matrix elements are explicitly denoted in (\ref{eq:a3})].

\section{Calculation of bremsstrahlung} 
\label{app:formula1}

The bremsstrahlung diagrams contain infrared divergences 
and the Feynman integrations require careful treatments, 
in particular, to calculate the 
angular dependent part. 
The terms in the first line of (\ref{eq:bremsmatrix}) do not contain 
${\mib p}_{\nu}$ and are angular-independent; hence, the angular integration 
of ${\mib k}$ is easy. 
Those in the second line of (\ref{eq:bremsmatrix}) are linear in 
${\mib p}_{\nu}$ and are proportional to ${\rm cos}\theta $. 
After some manipulations, (\ref{eq:brems1}) becomes

\begin{eqnarray} 
& &\hskip-1cm 
\frac{d\sigma (\bar \nu _{e} + p \longrightarrow e^{+} + n + \gamma )} 
{d({\rm cos}\theta )} 
\nonumber \\ 
& &= 
\frac{G_{V}^{2}}{\pi }\left ( \frac{e^{2}}{4\pi ^{2}}\right ) 
\Bigg [ 
(f_{V}^{2}+3g_{A}^{2}) \Bigg \{ -E^{2} {\rm tanh}^{-1}\beta \cdot 
{\rm log}\left ( \frac{\lambda }{E-m_{e}}\right ) +\sum _{k=1}^{6} 
{\cal I}_{k} \Bigg \} 
\nonumber \\ 
& & \hskip2.5cm +(f_{V}^{2} - g_{A}^{2}) 
\sum _{k=1}^{5}{\cal J}_{k} \Bigg ]. 
\end{eqnarray} 
Here the integrals ${\cal I}_{k}$, which do not contain 
$\cos\theta$, are given by 
\begin{eqnarray} 
{\cal I}_{1}&\equiv & E^{2}\int _{m_{e}}^{E-\lambda} dE' \frac{1}{E-E'} 
{\rm log}\left (\frac{E'+\vert {\mib \ell '}\vert}{ 
E(1+\beta )} \right ) 
\nonumber \\ 
&=& E^{2} \left \{ 
L\left (\frac{2\beta }{1+\beta }\right )+{\rm tanh}^{-1}\beta 
\cdot {\rm log}\left (\frac{2(E+m_{e})}{m_{e}}\right ) 
-({\rm tanh}^{-1}\beta )^{2}\right \}, 
\nonumber \\
\label{eq:cali1} 
\\ 
{\cal I}_{2}&\equiv & -E\int _{m_{e}}^{E-\lambda} dE' {\rm log} 
\left (\frac{E'+\vert {\mib \ell '}\vert}{m_{e}} \right ) 
\nonumber \\ 
&=& E^{2} \left ( -{\rm tanh}^{-1}\beta  + \beta  \right ), 
\\ 
{\cal I}_{3}&\equiv & \frac{1}{2} \int _{m_{e}}^{E-\lambda} dE' 
(E-E'){\rm log} 
\left (\frac{E'+\vert {\mib \ell '}\vert}{m_{e}} \right ) 
\nonumber \\ 
&=&E^{2}\left ( 
\frac{3-\beta ^{2}}{8}{\rm tanh}^{-1}\beta - 
\frac{3 \beta }{8}\right ), \\ 
{\cal I}_{4}&\equiv & 
-\frac{E}{2} \int _{m_{e}}^{E-\lambda }dE' 
\frac{1}{(E-E')^{2}}\vert {\mib k}\vert \vert {\mib \ell}' \vert 
\nonumber \\ 
&=& 
E^{2}\left \{ 
\frac{\beta }{2}{\rm log}\left ( 
\frac{\lambda }{m_{e}}\frac{1-\beta ^{2}}{4\beta ^{2}}\right ) 
+\frac{1}{2} {\rm tanh}^{-1}\beta +\beta \right \} , 
\\ 
{\cal I}_{5}&\equiv & 
\frac{1}{2}m_{e}^{2}E \int _{m_{e}}^{E-\lambda }dE' 
\Bigg \{ 
\frac{1}{2E'(E-E')+2\vert {\mib k}\vert \vert {\mib \ell }\vert 
+\lambda ^{2}} 
\nonumber \\
& & \hskip3cm  - 
\frac{1}{2E'(E-E')-2\vert {\mib k}\vert \vert {\mib \ell }\vert 
+\lambda ^{2}}\Bigg \} 
\nonumber \\ 
&=&E^{2} \left \{ 
\frac{\beta}{2} + \frac{\beta }{2}{\rm log}\left ( 
\frac{1-\beta ^{2}}{4\beta ^{2}}\right ) + {\rm tanh}^{-1}\beta 
+\frac{\beta }{2}{\rm log}\left ( 
\frac{\lambda }{m_{e}}\right )\right \} , 
\\ 
{\cal I}_{6}&\equiv &\frac{E^{2}}{2}\int _{m_{e}}^{E-\lambda } 
dE' \frac{1}{E-E'}{\rm log}\left ( 
\frac{2E'(E-E')+2\vert {\mib k}\vert \vert {\mib \ell }\vert 
+\lambda ^{2}}{2E'(E-E')-2\vert {\mib k}\vert \vert {\mib \ell } 
\vert +\lambda ^{2}}\cdot \frac{E'-\vert {\mib \ell '} \vert } 
{E'+ \vert {\mib \ell '} \vert} 
\right ) 
\nonumber \\ 
&=& 
E^{2}\left \{ 
\frac{1}{2}L\left ( \frac{2\beta }{1+\beta }\right )-\frac{1}{2} 
({\rm tanh}^{-1}\beta )^{2}+{\rm tanh}^{-1}\beta \cdot {\rm log}2 
\right \}. 
\label{eq:cali6} 
\end{eqnarray} 
Collecting (\ref{eq:cali1}) to (\ref{eq:cali6}) together, 
we obtain the angular independent part 
(\ref{eq:gb1beta}). 

The integrals in (\ref{eq:bremsmatrix}) that are proportional to 
$\cos\theta$ are given by 
\begin{eqnarray} 
{\cal J}_{1}&\equiv & 
\frac{1}{2\pi E_{\nu}} 
\int _{m_{e}}^{E-\lambda } 
dE' \vert {\mib \ell }' \vert \int \frac{d^{3}{\mib k}}{\omega} 
\delta (E-E'-\omega) 
\frac{({\mib \ell '}\cdot {\mib p}_{\nu})}{(2k\cdot \ell '+ 
\lambda ^{2})^{2}}\left \{ 
{\mib \ell '}^{2}-\frac{({\mib k}\cdot {\mib \ell '})^{2}}{ 
\omega ^{2}} 
\right \} 
\nonumber \\ 
&=&E^{2}\beta {\rm cos}\theta \Bigg \{ \frac{2}{\beta }+\frac{5\beta }{4}- 
\frac{2\sqrt{1-\beta ^{2}}}{\beta }-(1+\beta ){\rm tanh}^{-1}\beta 
\nonumber \\
& & 
+\left ( 
-\frac{1}{4\beta }-\frac{\beta }{4}+1\right ) ({\rm tanh}^{-1}\beta )^{2} 
-\frac{1}{2}L\left ( \frac{2\beta }{1+\beta } \right ) + 
2L\left (1-\sqrt{\frac{1-\beta }{1+\beta }} \right ) 
\nonumber \\ 
& & + 
(\beta -{\rm tanh}^{-1}\beta ){\rm log}\left ( 
\frac{\lambda }{2m_{e}}\left (1+\frac{1}{\beta }\right ) 
\frac{\sqrt{1+\beta }+\sqrt{1-\beta }}{\sqrt{1+\beta }-\sqrt{1-\beta }} 
\right ) \Bigg \} , 
\label{eq:C8} 
\\ 
{\cal J}_{2}&\equiv & 
\frac{1}{2\pi E_{\nu}} 
\int _{m_{e}}^{E-\lambda } 
dE' \vert {\mib \ell }' \vert 
\int \frac{d^{3}{\mib k}}{ 
\omega}\delta (E-E'-\omega)\frac{({\mib k}\cdot {\mib p}_{\nu})}{(2k\cdot 
\ell '+\lambda ^{2})^{2}}\left \{ 
{\mib \ell '}^{2}-\frac{({\mib k}\cdot {\mib \ell '})^{2}}{ 
\omega ^{2}}\right \} 
\nonumber \\ 
&=& 
E^{2}\beta  {\rm cos}\theta \Bigg \{ 
\frac{1}{2}{\rm tanh}^{-1}\beta - \beta 
+\frac{1-\beta ^{2} }{2\beta }({\rm tanh}^{-1}\beta )^{2} 
\Bigg \}, 
\\ 
{\cal J}_{3}&\equiv & 
\frac{1}{2\pi E_{\nu}} 
\int _{m_{e}}^{E-\lambda } 
dE' \vert {\mib \ell }' \vert 
\int \frac{d^{3}{\mib k}}{ 
\omega}\delta (E-E'-\omega)\frac{(k\cdot \ell ')({\mib k} \cdot 
{\mib p}_{\nu})}{(2k\cdot \ell '+ 
\lambda ^{2})^{2}} 
\nonumber \\ 
&=& 
E^{2}\beta {\rm cos}\theta \left \{ 
\frac{1}{4}{\rm tanh}^{-1}\beta -\frac{1-\beta ^{2}}{8\beta } 
({\rm tanh}^{-1}\beta )^{2}-\frac{1}{\beta }+\frac{3\beta }{8} 
+\frac{\sqrt{1-\beta ^{2}}}{\beta } 
\right \} , 
\nonumber \\
\\ 
{\cal J}_{4}&\equiv & 
\frac{1}{2\pi E_{\nu}} 
\int _{m_{e}}^{E-\lambda } 
dE' \vert {\mib \ell }' \vert 
\int \frac{d^{3}{\mib k}}{ 
\omega}\delta (E-E'-\omega)\frac{ 
(k\cdot \ell ')({\mib \ell '} \cdot {\mib p}_{\nu}) 
}{(2k\cdot \ell '+ 
\lambda ^{2})^{2}} 
\nonumber \\ 
&=& E^{2}\beta {\rm cos}\theta \left \{ 
\frac{1}{4}{\rm tanh}^{-1}\beta-\frac{1-\beta ^{2}}{8\beta } 
({\rm tanh}^{-1} \beta )^{2} 
-\frac{\beta }{8} 
\right \}, 
\\ 
{\cal J}_{5}&\equiv & 
-\frac{1}{2\pi E_{\nu}} 
\int _{m_{e}}^{E-\lambda } 
dE' \vert {\mib \ell }' \vert 
\int \frac{d^{3}{\mib k}}{ 
\omega}\delta (E-E'-\omega)\frac{ 
(k\cdot \ell ') 
}{(2k\cdot \ell '+ 
\lambda ^{2})^{2}} 
\frac{({\mib k}\cdot {\mib \ell '}) 
({\mib k} \cdot {\mib p}_{\nu})}{\omega ^{2}} 
\nonumber \\ 
&=& 
E^{2}\beta {\rm cos}\theta 
\left \{ 
-\frac{1}{4}{\rm tanh}^{-1}\beta 
-\frac{1-\beta ^{2}}{8\beta }({\rm tanh}^{-1}\beta )^{2}+\frac{3\beta }{8} 
\right \}. 
\end{eqnarray} 
Collecting these terms, we get (\ref{eq:new1}). 

Let us now explain the treatment of infrared divergences that appear 
in the integral of ${\cal J}_k$,  taking ${\cal J}_{1}$ as an example. 
After angular integration of the ${\mib k}$ variable, the integral is 
\begin{eqnarray} 
{\cal J}_{1}&=& 
{\rm cos }\theta 
\int _{m_{e}}^{E-\lambda} dE' \vert {\mib \ell }' \vert 
\Bigg [ 
-\frac{m_{e}^{2}}{2} 
\bigg  \{ 
\frac{1}{2E'(E-E')-2\vert {\mib k}\vert \vert 
{\mib \ell }'\vert +\lambda ^{2}} 
\nonumber \\
& & 
- \frac{1}{2E'(E-E')+2\vert {\mib k}\vert \vert 
{\mib \ell }'\vert +\lambda ^{2}} 
\bigg \} 
\nonumber \\ 
& & -\frac{ 
\vert {\mib k}\vert \vert {\mib \ell }'\vert }{ 
2(E-E')^{2}} 
\nonumber \\
& & - 
\frac{E'}{2(E-E')}{\rm log } 
\left ( 
\frac{2E'(E-E')-2\vert {\mib k}\vert \vert 
{\mib \ell }'\vert +\lambda ^{2}}{ 
2E'(E-E')+2\vert {\mib k}\vert \vert 
{\mib \ell }'\vert +\lambda ^{2}} 
\right ) 
\Bigg ]. 
\label{eq:j1anew} 
\end{eqnarray} 
A  special care is needed because 
there are two sources of the photon mass ($\lambda $) 
dependence. One comes from the the integrand and the other 
 from the integration region. To handle this problem, 
we split  the integral into   two terms, as 
${\cal J}_{1}={\cal J}_{1}'+({\cal J}_{1}-{\cal J}_{1}')$, 
where 
\begin{eqnarray} 
{\cal J}_{1}' \equiv 
{\rm cos}\theta \int_{m_{e}}^{E-\lambda} dE' 
\vert {\mib \ell }'\vert \Bigg [ 
-\frac{\vert {\mib \ell }'\vert }{E-E'} 
-\frac{E'}{2(E-E')}{\rm log}\left (\frac{ 
E'-\vert {\mib \ell}'\vert }{E'+\vert {\mib \ell}'\vert } 
\right )\Bigg ]\ , 
\nonumber \\ 
\label{eq:j1dashanew} 
\end{eqnarray} 
which is defined by putting $\lambda =0$ in the 
integrand of (\ref{eq:j1anew}), while retaining  $\lambda $ that appears 
in the integration region of $E'$.  (Recall the relation 
$\vert {\mib \ell }'\vert =\sqrt{E'^{2}-m_{e}^{2}}$ 
and 
$\vert {\mib k}\vert =\sqrt{(E-E')^{2}-\lambda ^{2}}$) . 

Integral (\ref{eq:j1dashanew}) is readily handled by changing the 
integration variable as 
\begin{eqnarray} 
E'\equiv \frac{m_{e}}{2}\left (\xi +\frac{1}{\xi}\right ), 
\end{eqnarray} 
with the result, 
\begin{eqnarray} 
{\cal J}_{1}'&=& 
E^{2}\beta {\rm cos}\theta \Bigg [ 
\frac{3}{4}\beta + \frac{2}{\beta} 
-\frac{2\sqrt{1-\beta ^{2}}}{\beta} 
\nonumber \\ 
& & 
-\left ( \frac{3}{2}+\beta \right ){\rm tanh}^{-1}\beta 
+\left (-\frac{1}{4\beta} - \frac{\beta }{4} +\frac{3}{2} 
\right )({\rm tanh}^{-1}\beta )^{2} 
\nonumber \\ 
& & 
-L\left ( \frac{2\beta }{1+\beta }\right ) 
+2L\left ( 1-\sqrt{\frac{1-\beta }{1+\beta }}\right ) 
\nonumber \\ 
& & + \left (\beta -{\rm tanh}^{-1}\beta \right ) 
{\rm log}\left (\frac{\lambda }{m_{e}}\left (1+\frac{1}{\beta } 
\right )\frac{\sqrt{1+\beta}+\sqrt{1-\beta }}{ 
\sqrt{1+\beta}- \sqrt{1-\beta }} 
\right ) 
\Bigg ]. 
\nonumber \\
\label{eq:j1dashb} 
\end{eqnarray} 

The evaluation of ${\cal J}_{1}-{\cal J}_{1}'$ requires 
a great care, since its integrand is non-vanishing 
only in the vicinity of the edge point $E'=E-\lambda $. 
To perform this integration, we choose the 
variable, 
\begin{eqnarray} 
\eta =\frac{E-E'+\sqrt{(E-E')^{2}-\lambda ^{2}}}{\lambda}. 
\end{eqnarray} 
In the limit $\lambda \longrightarrow 0$, \begin{eqnarray} 
{\cal J}_{1}-{\cal J}_{1}' 
&=& E^{2} \beta {\rm cos}\theta 
\Bigg [ 
-\frac{1}{2}({\rm tanh}^{-1} \beta )^{2}+ 
\frac{\beta }{2}+\frac{1}{2}{\rm tanh}^{-1}\beta \nonumber \\ 
& & 
- \left (\beta -{\rm tanh}^{-1}\beta \right ){\rm log}2 
+\frac{1}{2}L\left ( \frac{2\beta }{1+\beta } 
\right ) 
\Bigg ]\, 
\label{eq:j1minusjidash} 
\end{eqnarray} 
which is infrared finite. By adding (\ref{eq:j1minusjidash}) and 
(\ref{eq:j1dashb}), the integration of ${\cal J}_{1}$ is completed 
to give (\ref{eq:C8}). 

\section{Currents and their commutation relations} 
\label{app:ca}

We present explicitly the currents and the current commutation relations 
that appear in Sect. \ref{sec:universality}. 
Consider a fermion field $\psi $, which belongs to some representation 
of the  internal symmetry group and satisfies the equal time (ET) 
canonical commutation relations 
$ 
\left \{ 
\psi _{\alpha }(x), \psi _{\beta }^{\dag}(y) 
\right \}_{\rm ET}= 
\delta _{\alpha \beta}\delta ^{3}({\mib x}-{\mib y}) 
$. 
The vector and axial-vector currents are constructed as, 
\begin{eqnarray} 
j_{\mu}^{a}(x)=\bar \psi (x)\gamma _{\mu}T^{a}\psi (x), \hskip0.5cm 
j_{\mu}^{5a}(x)=\bar \psi (x)\gamma _{\mu}\gamma ^{5}T^{a}\psi (x), 
\end{eqnarray} 
where $T^{a}$ are the generators of the internal symmetry group. 

By applying the canonical commutation relations, 
we obtain 
\begin{eqnarray} 
\left [ 
j_{\mu}^{a}(x), j_{\nu}^{b}(y) 
\right ]_{\rm ET}  
&=& \bar \psi (x)\gamma _{\mu}\gamma ^{0} 
\gamma _{\nu}T^{a}T^{b}\psi (y)\delta ^{3}({\mib x}-{\mib y}) 
\nonumber \\
& & 
- \bar \psi (y)\gamma _{\nu}\gamma ^{0} 
\gamma _{\mu}T^{b}T^{a}\psi (x)\delta ^{3}({\mib x}-{\mib y}), 
\label{eq:vv} 
\\ 
\left [ 
j_{\mu}^{5a}(x), j_{\nu}^{b}(y) 
\right ]_{\rm ET} 
&=& \bar \psi (x)\gamma _{\mu}\gamma ^{5}\gamma ^{0} 
\gamma _{\nu}T^{a}T^{b}\psi (y)\delta ^{3}({\mib x}-{\mib y}) 
\nonumber \\
& &- \bar \psi (y)\gamma _{\nu}\gamma ^{0} 
\gamma _{\mu}\gamma ^{5}T^{b}T^{a}\psi (x)\delta ^{3}({\mib x}-{\mib y}), 
\label{eq:av} 
\\ 
\left [ 
j_{\mu}^{5a}(x), j_{\nu}^{5b}(y) 
\right ]_{\rm ET}  
&=& \bar \psi (x)\gamma _{\mu}\gamma ^{0} 
\gamma _{\nu}T^{a}T^{b}\psi (y)\delta ^{3}({\mib x}-{\mib y}) 
\nonumber \\
& &- \bar \psi (y)\gamma _{\nu}\gamma ^{0} 
\gamma _{\mu}T^{b}T^{a}\psi (x)\delta ^{3}({\mib x}-{\mib y}). 
\label{eq:aa} 
\end{eqnarray} 
 From (\ref{eq:vv})-(\ref{eq:aa}), conventional 
current algebra follows for $\nu =0$. 

The commutation relations we use are 
derived from (\ref{eq:vv})-(\ref{eq:aa}). 
One class of the commutation relations is 
\begin{eqnarray} 
g^{\mu \nu}\left [ 
j_{\mu}^{a}(x),  j_{\nu}^{b}(y) 
\right ]_{\rm ET} 
&=& 
-2 \bar \psi (x) \gamma _{0}\left [ 
T^{a}, T^{b} \right ]\psi (x)\delta ^{3}({\mib x}-{\mib y}),  
\\ 
g^{\mu \nu}\left [ 
j_{\mu}^{5a}(x),  j_{\nu}^{b}(y) 
\right ]_{\rm ET} 
&=& 
-2 \bar \psi (x) \gamma _{0} \gamma ^{5} \left [ 
T^{a}, T^{b} \right ]\psi (x)\delta ^{3}({\mib x}-{\mib y}) , 
\\ 
g^{\mu \nu}\left [ 
j_{\mu}^{5a}(x),  j_{\nu}^{5b}(y) 
\right ]_{\rm ET} 
&=& 
-2 \bar \psi (x) \gamma _{0}\left [ 
T^{a}, T^{b} \right ]\psi (x)\delta ^{3}({\mib x}-{\mib y}). 
\end{eqnarray} 
For the other class of the commutation relations, 
by noting 
\begin{eqnarray} 
\varepsilon ^{\lambda \mu   \nu \rho } 
\gamma _{\mu }\gamma _{0 }\gamma _{\nu} = 
2i\left ( 
g^{\lambda }_{0}g^{\rho }_{\sigma}-g^{\lambda}_{\sigma}g^{\rho}_{0} 
\right )\gamma ^{\sigma }\gamma ^{5}. 
\end{eqnarray} 
we obtain 
\begin{eqnarray} 
& & \hskip-1cm \varepsilon^{\lambda \mu \nu \rho} \left [ 
j_{\mu}^{a}(x),  j_{\nu}^{b}(y) 
\right ]_{\rm ET} 
\nonumber \\
& & = 
2i  \left ( 
g^{\lambda }_{0}g^{\rho }_{\sigma}-g^{\lambda}_{\sigma}g^{\rho}_{0} 
\right ) 
\bar \psi (x) \gamma ^{\sigma }\gamma ^{5} \left \{ 
T^{a}, T^{b} \right \} \psi (x)\delta ^{3}({\mib x}-{\mib y}), 
\\ 
& & \hskip-1cm \varepsilon^{\lambda \mu \nu \rho}  \left [ 
j_{\mu}^{5a}(x),  j_{\nu}^{b}(y) 
\right ]_{\rm ET} 
\nonumber \\
& & = 
2i  \left ( 
g^{\lambda }_{0}g^{\rho }_{\sigma}-g^{\lambda}_{\sigma}g^{\rho}_{0} 
\right ) 
\bar \psi (x) \gamma ^{\sigma } \left \{ 
T^{a}, T^{b} \right \} \psi (x)\delta ^{3}({\mib x}-{\mib y}), 
\\ 
& & \hskip-1cm \varepsilon^{\lambda \mu \nu \rho}  \left [ 
j_{\mu}^{5a}(x),  j_{\nu}^{5b}(y) 
\right ]_{\rm ET} 
\nonumber \\ 
& & = 
2i  \left ( 
g^{\lambda }_{0}g^{\rho }_{\sigma}-g^{\lambda}_{\sigma}g^{\rho}_{0} 
\right ) 
\bar \psi (x) \gamma ^{\sigma }\gamma ^{5} \left \{ 
T^{a}, T^{b} \right \} \psi (x)\delta ^{3}({\mib x}-{\mib y}). 
\end{eqnarray}

Let us consider the case of $SU(2)$ and suppose 
that  $\psi $ belongs to the doublet representation. 
The electromagnetic and weak currents are \begin{eqnarray} 
j_{\mu }^{\rm em}=\bar \psi \gamma _{\mu}T^{Q} \psi , 
\hskip0.5cm 
t_{\lambda } = \bar \psi \gamma _{\lambda }  T^{-} \psi 
- \bar \psi \gamma _{\lambda }\gamma ^{5}T^{-} \psi , 
\end{eqnarray} 
where the charge matrix $T^{Q}$ and $T^{-}$ are defined by 
\begin{eqnarray} 
& &T^{Q}=\left ( 
\begin{tabular}{c c} 
$Q_{+}$ & 0 \\ 
0 & $Q_{-}$ 
\end{tabular} 
\right ), 
\hskip0.5cm 
T^{-}=\left ( 
\begin{tabular}{c c} 
0 & 0 \\ 
1 & 0 
\end{tabular} 
\right );
\end{eqnarray} 
the electric charges of upper and lower components of the doublet 
are denoted by $Q_{+}$ and $Q_{-}$. 

The commutation and anticommutation relations between 
$T^{Q}$ and $T^{-}$ are 
\begin{eqnarray} 
& &\left [ 
T^{-}, T^{Q} 
\right ]=(Q_{+}-Q_{-})T^{-}=T^{-}, 
\\ 
& &\left \{ 
T^{-}, T^{Q} 
\right \}=(Q_{+}+Q_{-}) T^{-}\equiv 2\bar Q T^{-}. 
\end{eqnarray} 
The relations are translated as \begin{eqnarray} 
g^{\lambda \mu} \left [ 
t_{\lambda }(x), j_{\mu }^{\rm em}(y) 
\right ]_{\rm ET}&=&-2t_{0}(x)\delta ^{3}({\mib x}-{\mib y}), 
\label{eq:gdecontract} 
\\ 
\varepsilon^{ \lambda \mu \nu \rho } \left [ 
t_{\mu }(x), j_{\nu }^{\rm em}(y) 
\right ]_{\rm ET} 
&=& 
-4i\bar Q 
\left ( 
g^{\lambda}_{0}g^{\rho}_{\sigma}-g^{\lambda}_{\sigma}g^{\rho}_{0} 
\right ) 
 t^{\sigma}(x) 
\delta ^{3}({\mib x}-{\mib y}).
\nonumber \\
\label{eq:epsilondecontract} 
\end{eqnarray} 
The relation $Q_{+}-Q_{-}=1$ holds independently of the model 
of the constituents of hadrons; hence (\ref{eq:gdecontract}) 
is model-independent. 
On the contrary, the mean charge $\bar Q$ 
depends on the constituents, e.g., whether $\psi $ denotes 
the nucleon- or quark-doublet ($\bar Q=1/2$ or 1/6), 
so (\ref{eq:epsilondecontract}) is  model-dependent. 

Some commutators of the local currents may in principle receive 
model-dependent Schwinger terms. 
Within the standard model, however, the Schwinger term is c-numbers,
and it does not contribute to beta decay of nucleons,
as argued by Sirlin \cite{sirlinrev}.

\section{Calculation of ${\cal M}^{(v3,VA)}$ and ${\cal M}^{(v3,wm)}$} 
\label{app:inner}

To shuffle the gamma matrices in  (\ref{eq:pv3va}), 
we use 
\begin{eqnarray} 
& & \hskip-1cm \frac{1}{2} 
\left \{ 
\gamma _{\lambda} \left ( 
f_{V}F_{V}(k^{2})-g_{A}F_{A}(k^{2})\gamma ^{5} \right ), 
\sigma _{\mu \rho}\right \}
\nonumber \\
& & = 
-\varepsilon _{\lambda \mu \rho \sigma }\gamma ^{\sigma }\gamma ^{5} 
\left ( f_{V}F_{V}(k^{2})-g_{A}F_{A}(k^{2})\gamma ^{5} \right ), 
\\ 
& &  \hskip-1cm \frac{1}{2} 
\left [ 
\gamma _{\lambda} \left ( 
f_{V}F_{V}(k^{2})-g_{A}F_{A}(k^{2})\gamma ^{5} \right ), 
\sigma _{\mu \rho}\right ]
\nonumber \\
& & = i\left ( 
g_{\lambda \mu}\gamma _{\rho}-g_{\lambda \rho} \gamma _{\mu} 
\right ) 
\left ( f_{V}F_{V}(k^{2})-g_{A}F_{A}(k^{2})\gamma ^{5} \right ), 
\end{eqnarray} 
and simplify nucleon's gamma matrices in (\ref{eq:pv3va}) using 
\begin{eqnarray} 
& & \hskip-1.5cm \gamma _{\lambda} \left ( 
f_{V}F_{V}(k^{2})-g_{A}F_{A}(k^{2})\gamma ^{5} \right ) 
\sigma _{\mu \rho}
\nonumber \\
&=& \frac{1}{2}\left \{ 
\gamma _{\lambda} \left ( 
f_{V}F_{V}(k^{2})-g_{A}F_{A}(k^{2})\gamma ^{5} \right ), 
\sigma _{\mu \rho}\right \} 
\nonumber \\
& &  +\frac{1}{2}\left [\gamma _{\lambda} \left ( 
f_{V}F_{V}(k^{2})-g_{A}F_{A}(k^{2})\gamma ^{5} \right ), 
\sigma _{\mu \rho}\right ] . 
\end{eqnarray} 
Applying these formulae and using the definition of 
${\cal C}_{\sigma,\tau}$ in 
(\ref{eq:formnum1}) and (\ref{eq:formnum2}), (\ref{eq:pv3va}) 
is re-expressed as \begin{eqnarray} 
{\cal M}_{p}^{(v3,VA)}&=&\frac{G_{V}}{\sqrt{2}} 
\frac{e^{2}}{16\pi ^{2}}\bar v _{\nu}\gamma ^{\lambda } 
(1-\gamma ^{5})\sigma ^{\mu \nu}v_{e}(\ell) 
\nonumber \\ 
& &   \hskip-1.5cm 
  \times \Bigg [\varepsilon _{\lambda \mu \nu \sigma }\bar u 
_{n}(p_{1})\gamma 
^{\sigma }\gamma ^{5}\left (f_{V}{\cal C}_\sigma^{(p,V)}-g_{A}\gamma ^{5} 
{\cal C}_\sigma^{(p, A)}\right )u_{p}(p_{2}) 
\nonumber \\ 
& &  \hskip-1.5cm   
+\frac{1}{m_{p}^{2}}\varepsilon _{\lambda \mu \rho \sigma } 
(p_{2})_{\nu}(p_{2})^{\rho}\bar u _{n}(p_{1})\gamma ^{\sigma } 
\gamma ^{5}\left ( 
f_{V}{\cal C}_\tau^{(p,V)}-g_{A}\gamma ^{5}{\cal C}_\tau^{(p,A)} 
\right )u_{p}(p_{2}) 
\nonumber \\ 
& &    \hskip-1.5cm   
-i \bar u _{n}(p_{1}) 
\left ( g_{\lambda \mu}\gamma _{\nu}-g_{\lambda \nu}\gamma _{\mu} 
\right ) \left (f_{V}{\cal C}_\sigma^{(p,V)}-g_{A}\gamma ^{5} 
{\cal C}_\sigma^{(p, A)}\right )u_{p}(p_{2}) 
\nonumber \\ 
& &  \hskip-1.5cm 
-\frac{i}{m_{p}^{2}} 
(p_{2})^{\nu}(p_{2})^{\rho}\bar u _{n}(p_{1}) 
\left ( g_{\lambda \mu}\gamma _{\rho}-g_{\lambda \rho}\gamma _{\mu} 
\right ) 
\left ( 
f_{V}{\cal C}_\tau^{(p,V)}-g_{A}\gamma ^{5}{\cal C}_\tau^{(p, A)} 
\right )u_{p}(p_{2})\Bigg ]. 
\nonumber \\ 
\label{eq:appmpv3va} 
\end{eqnarray} 

We are interested only in those terms that 
are proportional to $f_{V}g_{A}$ in 
the amplitude squared. 
By inspecting the gamma matrices in (\ref{eq:appmpv3va}) in the 
nucleon static limit, we 
see that only the first and second terms in the 
brackets of (\ref{eq:appmpv3va}) are potential sources 
of the $f_{V}g_{A}$ terms; the third and fourth terms give 
those proportional to $f_{V}^{2}$ or $g_{A}^{2}$. 
The spin summation relevant to the first and 
second terms are 
\begin{eqnarray} 
& &\sum _{\rm spin}\varepsilon _{\lambda \mu \nu \sigma } 
[\bar u_{n}(p_{1})\gamma ^{\sigma }\gamma ^{5}\left ( 
f_{V}{\cal C}_{\sigma }^{(p, V)}-g_{A}\gamma ^{5} 
{\cal C}_{\sigma }^{(p, A)} 
\right )u_{p}(p_{2})] 
\nonumber \\
& & \hskip0.5cm \times 
[\bar u_{n}(p_{1})\gamma _{\rho}\left ( 
f_{V}-g_{A}\gamma ^{5}\right )u_{p}(p_{2})]^{*} 
\nonumber \\ 
& & \hskip1cm \times 
\sum _{\rm spin} 
[\bar v_{\nu}\gamma ^{\lambda }\sigma ^{\mu \nu}(1-\gamma ^{5})] 
[\bar v_{\nu}\gamma ^{\rho}(1-\gamma ^{5})v_{e}(\ell )]^{*} 
\nonumber  \\ 
& &= 384 m_{n}m_{p}E E_{\nu}f_{V}g_{A} 
\left \{ 
{\cal C}_{\sigma }^{(p, A)}(1+\beta {\rm cos}\theta ) + 
{\cal C}_{\sigma }^{(p, V)}(3-\beta {\rm cos }\theta ) 
\right \}, 
\\ 
& &\sum _{\rm spin}\frac{1}{m_{p}^{2}} 
\varepsilon _{\lambda \mu \tau \sigma } 
(p_{2})_{\nu}(p_{2})^{\tau } 
[\bar u_{n}(p_{1})\gamma ^{\sigma }\gamma ^{5}\left ( 
f_{V}{\cal C}_{\tau }^{(p, V)}-g_{A}\gamma ^{5} 
{\cal C}_{\tau }^{(p, A)} 
\right )u_{p}(p_{2})] 
\nonumber \\ 
& & 
\hskip0.5cm \times 
[\bar u_{n}(p_{1})\gamma _{\rho}\left ( 
f_{V}-g_{A}\gamma ^{5}\right )u_{p}(p_{2})]^{*} 
\nonumber \\ 
& & \hskip1cm \times \sum _{\rm spin} 
[\bar v_{\nu}\gamma ^{\lambda }\sigma ^{\mu \nu}(1-\gamma ^{5})] 
[\bar v_{\nu}\gamma ^{\rho}(1-\gamma ^{5})v_{e}(\ell )]^{*} 
\nonumber  \\ 
& &= 128 m_{n}m_{p}E E_{\nu}f_{V}g_{A} 
{\cal C}_{\tau }^{(p, V)}(3-\beta {\rm cos }\theta ). 
\end{eqnarray} 
It is obvious that $(1+\beta {\rm cos}\theta )$ and 
$(3-\beta {\rm cos}\theta )$ terms correspond 
to the Fermi and Gamow-Teller parts, respectively. 

The interference with ${\cal M}^{(0)}$ turns out to be 
\begin{eqnarray} 
& & \hskip-2cm \sum _{\rm spin}\left ( 
{\cal M}_{p}^{(v3, VA)}{\cal M}^{(0)*}+ 
{\cal M}_{p}^{(v3, VA)*}{\cal M}^{(0)} 
\right )\Bigg \vert _{f_{V}g_{A}} 
\nonumber \\ 
&=& 
32G_{V}^{2}m_{n}m_{p}EE_{\nu}\left ( \frac{e^{2}}{8\pi ^{2}} 
\right ) f_{V}g_{A}\Bigg [ 
6{\cal C}_{\sigma }^{(p, A)}(1+\beta {\rm cos}\theta )\langle 
1 \rangle ^{2} 
\nonumber \\ 
& &+(6{\cal C}_{\sigma }^{(p, V)}+2{\cal C}_{\tau }^{(p, V)}) 
(3-\beta {\rm cos}\theta )\frac{1}{3}\langle 
{\mib \sigma }\rangle ^{2} \Bigg ], 
\end{eqnarray} 
where only the $f_{V}g_{A}$ terms are retained. 
The same procedure applies to the neutron amplitude, 
and we arrive at (\ref{eq:syuusei}). 

For the weak magnetism terms, the gamma matrices in the last line of 
(\ref{eq:pv3wm}) are simplified by applying 
\begin{eqnarray} 
\left ( \sigma _{\lambda \rho} k^{\rho }\right ) 
\left ( \sigma _{\mu \nu} k^{\nu}\right ) 
&=& 
\frac{1}{2}\left \{ 
\sigma _{\lambda \rho} k^{\rho }, \sigma _{\mu \nu} k^{\nu} 
\right \} 
+\frac{1}{2}\left [ 
\sigma _{\lambda \rho} k^{\rho }, \sigma _{\mu \nu} k^{\nu} 
\right ], 
\\ 
\left [ 
\sigma _{\lambda \rho} k^{\rho }, \sigma _{\mu \nu} k^{\nu} 
\right ] 
&=& 
-ik_{\lambda}\sigma _{\mu \rho}k^{\rho} 
+ik_{\mu}\sigma _{\lambda \rho}k^{\rho} 
-ik^{2}\sigma _{\mu \nu} 
+i\varepsilon _{\lambda \mu \nu \kappa }k^{\nu} 
(\gamma \cdot k) \gamma ^{\kappa }\gamma ^{5}, 
\nonumber \\
\label{eq:e7} 
\\ 
\left \{ 
\sigma _{\lambda \rho} k^{\rho }, \sigma _{\mu \nu} k^{\nu} 
\right \} 
&=&-2\left ( 
k_{\lambda }k_{\mu}-k^{2}g_{\mu \lambda } \right ). 
\label{eq:e8} 
\end{eqnarray} 
Note that (\ref{eq:e7}) and (\ref{eq:e8}) are antisymmetric and symmetric 
under the interchange of the indices $\lambda $ and $\mu $. 
We also simplify the gamma matrices in the leptonic sector in 
(\ref{eq:pv3wm}) by using 
\begin{eqnarray} 
i\gamma ^{\lambda} (1-\gamma ^{5}) \sigma ^{\mu \nu}k_{\nu}&=& 
\left \{ 
\frac{1}{2}\left (k^{\lambda }\gamma ^{\mu}+k^{\mu}\gamma ^{\lambda } 
\right )-g^{\lambda \mu}(\gamma \cdot k) 
\right \}(1-\gamma ^{5}) 
\nonumber \\ 
& &+\left \{ \frac{1}{2}\left ( k^{\lambda }\gamma ^{\mu}-k^{\mu}\gamma 
^{\lambda } 
\right ) -i\varepsilon ^{\lambda \mu \nu \rho }\gamma _{\rho}\gamma ^{5} 
k_{\nu}\right \}(1-\gamma ^{5}). 
\nonumber \\
\label{eq:e9} 
\end{eqnarray} 
The first (second) line is symmetric (antisymmetric) 
under the interchange of $\lambda $ and $\mu $. 
Neglecting $m_{e}$, $\ell $ and $\lambda$, and using (\ref{eq:e7}), 
(\ref{eq:e8}) and (\ref{eq:e9}), we simplify (\ref{eq:pv3wm}) as 
\begin{eqnarray} 
& & \hskip-1cm  {\cal M}_{p}^{(v3,wm)} 
\nonumber \\
&=& 
\frac{i}{\sqrt{2}}G_{V}e^{2}\left ( \frac{i}{2m_{N}} \right ) 
\int \frac{d^{4}k}{(2\pi )^{4}}\frac{1}{(k^{2})^{2}} 
\frac{1}{(p_{2}-k)^{2}-m_{p}^{2}} 
F_{1}^{(p)}(k^{2})F_{W}(k^{2}) 
\nonumber \\ 
& & 
\times \Bigg \{ 
-[\bar v_{\nu}\gamma ^{\lambda }(1-\gamma ^{5})v_{e}(\ell )] 
[\bar u_{n}(p_{1})\sigma _{\lambda \rho}k^{\rho}u_{p}(p_{2})] 
(k^{2}-2k\cdot p_{2}) 
\nonumber \\ 
& & 
+2[\bar v_{\nu}\gamma ^{\lambda }(1-\gamma ^{5})\left ( k\cdot p_{2}+ 
i\sigma ^{\mu \nu}k_{\nu}(p_{2})_{\mu} \right )v_{e}(\ell )] 
[\bar u_{n}(p_{1})\sigma _{\lambda \rho}k^{\rho}u_{p}(p_{2})] 
\Bigg \} 
\nonumber \\ 
& &+\frac{i}{\sqrt{2}}G_{V}e^{2}\left ( \frac{i}{2m_{N}} \right ) 
\left ( \frac{-i}{2} \right ) 
\int \frac{d^{4}k}{(2\pi )^{4}}\frac{1}{(k^{2})^{2}} 
\frac{1}{(p_{2}-k)^{2}-m_{p}^{2}} 
\nonumber \\ 
& &\times \left \{ F_{1}^{(p)}(k^{2}) + F_{2}^{(p)}(k^{2}) \right \} 
F_{W}(k^{2}) 
\nonumber \\ 
& &\times \Bigg \{ 
-6 [\bar v_{\nu }\gamma \cdot k (1-\gamma ^{5}) v_{e}(\ell )] 
[\bar u_{n}(p_{1})u_{p}(p_{2})] 
\nonumber \\ 
& &+ 2 \varepsilon ^{\lambda \mu \nu \rho} k_{\nu} [ \bar v_{\nu} 
\gamma _{\rho}(1-\gamma ^{5})v_{e}(\ell )] 
[\bar u_{n}(p_{1})\sigma _{\lambda \mu}u_{p}(p_{2})] 
\Bigg \}. 
\label{eq:3461} 
\end{eqnarray} 

In (\ref{eq:3461}) the term containing 
$[\bar u_{n}(p_{1})\sigma _{\lambda \rho}k^{\rho}u_{p}(p_{2})] 
(k^{2}-2k\cdot p_{2}) $ does not survive the symmetric integration 
over $k$. In the static nucleon limit, we retain 
only those terms containing 
$\bar u_{n}(p_{1})u_{p}(p_{2})$ for the Fermi part and 
$\bar u_{n}(p_{1})\sigma _{ij}u_{p}(p_{2})$ for the Gamow-Teller part. 
Using the definitions in (\ref{eq:dspdtp}) and (\ref{eq:e(p)}), 
(\ref{eq:3461}) is rewritten  
\begin{eqnarray} 
& & \hskip-1cm {\cal M}_{p}^{(v3, wm)}
\nonumber \\
&=& 
\frac{G_{V}}{\sqrt{2}}\frac{e^{2}}{16\pi ^{2}} 
\left (\frac{m_{p}}{m_{N}}\right ){\cal D}_\sigma^{(p)}\sum _{i,j=1}^{3} 
[\bar v_{\nu}\gamma ^{i}(1-\gamma ^{5})\sigma ^{0j}v_{e}(\ell )] 
[\bar u_{n}(p_{1})\sigma _{ij}u_{p}(p_{2})] 
\nonumber \\ 
& &+\frac{G_{V}}{\sqrt{2}}\frac{e^{2}}{16\pi ^{2}} 
\left (\frac{m_{p}}{m_{N}}\right ) 
{\cal E}^{(p)}\Bigg \{ 
\frac{3}{2}[\bar v_{\nu}\gamma ^{0}(1-\gamma ^{5})v_{e}(\ell )] 
[ \bar u_{n}(p_{1})u_{p}(p_{2})] 
\nonumber \\ 
& &-\frac{1}{2}\sum _{i,j,k=1}^{3}\varepsilon ^{0ijk} 
[\bar v_{\nu}\gamma _{k}(1-\gamma ^{5})v_{e}(\ell )] 
[\bar u_{n}(p_{1})\sigma _{ij}u_{p}(p_{2})] \Bigg \}. 
\end{eqnarray} 

The following 
spin summations are employed to calculate the interference with 
${\cal M}^{(0)}$: 
\begin{eqnarray} 
& &\sum _{\rm spin}[\bar v_{\nu}\gamma ^{i}(1-\gamma ^{5}) 
\sigma ^{0j}v_{e}(\ell )][\bar v _{\nu}\gamma _{\rho} 
(1-\gamma ^{5})v_{e}(\ell )]^{*} 
\nonumber \\ 
& & \hskip0.5cm \times 
\sum _{\rm spin}[\bar u_{n}(p_{1})\sigma _{ij}u_{p}(p_{2})] 
[\bar u_{n}(p_{1})\gamma ^{\rho}(f_{V}-g_{A}\gamma ^{5})u_{p}(p_{2})]^{*} 
\nonumber \\ 
& & \hskip1cm = 128g_{A}m_{n}m_{p}EE_{\nu}(3-\beta {\rm cos}\theta ), 
\\ 
& &\sum _{\rm spin}[\bar v_{\nu}\gamma ^{0}(1- 
\gamma ^{5})v_{e}(\ell )] 
[\bar v _{\nu}\gamma _{\rho} 
(1-\gamma ^{5})v_{e}(\ell )]^{*} 
\nonumber \\ 
& & \hskip0.5cm \times 
\sum _{\rm spin}[\bar u_{n}(p_{1})u_{p}(p_{2})] 
[\bar u_{n}(p_{1})\gamma ^{\rho}(f_{V}-g_{A}\gamma ^{5})u_{p}(p_{2})]^{*} 
\nonumber \\ 
& & \hskip1cm = 64f_{V}m_{n}m_{p}EE_{\nu}(1+\beta {\rm cos}\theta ), 
\\ 
& &\varepsilon ^{0ijk} \sum _{\rm spin} [\bar v_{\nu}\gamma _{k}(1- 
\gamma ^{5})v_{e}(\ell )] 
[\bar v _{\nu}\gamma _{\rho} 
(1-\gamma ^{5})v_{e}(\ell )]^{*} 
\nonumber \\ 
& & \hskip0.5cm \times 
\sum _{\rm spin}[\bar u_{n}(p_{1})\sigma _{ij} u_{p}(p_{2})] 
[\bar u_{n}(p_{1})\gamma ^{\rho}(f_{V}-g_{A}\gamma ^{5})u_{p}(p_{2})]^{*} 
\nonumber \\ 
& & \hskip1cm = -128 g_{A}m_{n}m_{p}EE_{\nu}(3-\beta {\rm cos}\theta ). 
\end{eqnarray} 
We obtain for ${\cal M}_{p}^{(v3,wm)}$ 
\begin{eqnarray} 
& & \hskip-0.5cm 
\sum _{\rm spin}\left \{ {\cal M}_{p}^{(v3,wm)}{\cal M}^{(0)*}+ 
{\cal M}_{p}^{(v3,wm)*}{\cal M}^{(0)}\right \} 
\nonumber \\ 
&=& 32G_{V}^{2}m_{n}m_{p}EE_{\nu}\left ( 
\frac{e^{2}}{8\pi ^{2}}\right ) 
\frac{1}{m_{N}}\Bigg [ 
2g_{A}m_{p}{\cal D}_{\sigma }^{(p)}(3-\beta {\rm cos}\theta) 
\nonumber \\
& & 
+g_{A}m_{p}{\cal E}^{(p)}(3-\beta {\rm cos}\theta ) 
+\frac{3}{2} f_{V}m_{p}{\cal E}^{(p)}(1+\beta {\rm cos}\theta ) 
\Bigg ]. 
\end{eqnarray} 
Similar calculations are made for ${\cal M}_{n}^{(v3, wm)}$ 
and we finally obtain (\ref{eq:2268}) 

\section{Nucleon Form Factors} 
\label{app:formfactor}

To calculate numerical constants that appear in (\ref{eq:formnum1}) 
and (\ref{eq:formnum2}), 
we use the dipole form factors ($\mu _{p}=$ 
1.793, $\mu _{n}=-1.913$): 
\begin{eqnarray} 
F_{1}^{(p)}(k^{2})+ 
F_{1}^{(n)}(k^{2}) 
=\frac{1}{\left (1-k^{2}/m _{V}^{2}\right )^{2}}, 
\\ 
F_{1}^{(p)}(k^{2})-F_{1}^{(n)}(k^{2}) 
=\frac{1}{\left (1-k^{2}/m _{V}^{2} \right )^{2}}, 
\\ 
F_{2}^{(p)}(k^{2})+F_{2}^{(n)}(k^{2}) 
=\frac{\mu _{p}+\mu _{n}}{\left (1-k^{2}/m _{V}^{2} \right )^{2}}, 
\\ 
F_{2}^{(p)}(k^{2})-F_{2}^{(n)}(k^{2}) 
=\frac{\mu _{p}-\mu _{n}}{\left (1-k^{2}/m _{V}^{2} \right )^{2}}, 
\end{eqnarray} 
where we adopt $m_{V}=0.84$ {\rm GeV}. 
We also adopt the dipole form factors 
\begin{eqnarray} 
F_{V}(k^{2})=\frac{1}{(1-k^{2}/m _{V}^{2})^{2}}, \hskip1cm 
F_{A}(k^{2})=\frac{1}{(1-k^{2}/m _{A}^{2})^{2}}, 
\end{eqnarray} 
for the vector and axial-vector vertices; 
we take $m_{A}=1.05$ {\rm GeV}. Weak magnetism should also be 
endowed with the form factor 
\begin{eqnarray} 
F_{W}(k^{2})=\frac{\mu _{p}-\mu _{n}}{(1-k^{2}/m _{V}^{2})^{2}}. 
\end{eqnarray}

The integrals of (\ref{eq:formnum1}) are  inverted as \begin{eqnarray} 
\frac{i}{16\pi ^{2}} 
{\cal C}_\sigma^{(p, V)}=\frac{1}{3} 
\int \frac{d^{4}k}{(2\pi )^{4}}\frac{1}{(k^{2})^{2}} 
\frac{1}{(p_{2}-k)^{2}-m_{p}^{2}} 
\left \{ k^{2}-\frac{1}{m_{p}^{2}} 
(k\cdot p_{2})^{2}\right \}& & 
\nonumber \\ 
\times \left \{ 
F_{1}^{(p)}(k^{2})+F_{2}^{(p)}(k^{2}) 
\right \}F_{V}(k^{2})\ , 
\\ 
\frac{i}{16\pi ^{2}} 
{\cal C}_\tau^{(p, V)}=-\frac{1}{3} 
\int \frac{d^{4}k}{(2\pi )^{4}}\frac{1}{(k^{2})^{2}} 
\frac{1}{(p_{2}-k)^{2}-m_{p}^{2}} 
\left \{ k^{2}-\frac{4}{m_{p}^{2}} 
(k\cdot p_{2})^{2}\right \}& & 
\nonumber \\ 
\times \left \{ 
F_{1}^{(p)}(k^{2})+F_{2}^{(p)}(k^{2}) 
\right \}F_{V}(k^{2})\ . 
\nonumber \\ 
\end{eqnarray} 
We use integral representations employing the Feynman trick 
to compute numerically the constants in 
({\ref{eq:formnum1}) and (\ref{eq:formnum2}), 
\begin{eqnarray} 
& & \int \frac{d^{4}k}{(2\pi )^{4}}\frac{1}{(k^{2})^{2}} 
\frac{1}{(p_{2}-k)^{2}-m_{p}^{2}}\frac{1}{(1-k^{2}/\Lambda _{0}^{2})^{2}} 
\frac{1}{(1-k^{2}/\Lambda _{1}^{2})^{2}} 
\nonumber \\ 
& &\hskip3cm \times 
\left \{ k^{2}-\frac{2\xi _{0}}{m_{p}^{2}} (k\cdot p)^{2}\right \} 
\nonumber \\ 
&=& 
\Lambda _{0}^{2}\Lambda _{1}^{2}f(\Lambda _{0}, \Lambda _{1}) 
+\frac{\Lambda _{0}^{2}\Lambda _{1}^{2}}{\Lambda _{1}^{2}-\Lambda _{0}^{2}} 
\left \{ \Lambda _{0}^{2}f(\Lambda _{0}, 0) 
-\Lambda _{1}^{2}f(0, \Lambda _{1}) 
\right \}\ , 
\end{eqnarray} 
where 
\begin{eqnarray} 
f(\Lambda _{0}, \Lambda _{1}) 
&=&\frac{i}{16\pi ^{2}}\int _{0}^{1}dx \int _{0}^{1-x} 
dy x(1-x-y) 
\nonumber \\
& & \times 
\Bigg [ 
\frac{(2-\xi _{0})}{\left \{ 
y^{2}m_{p}^{2}+x\Lambda _{0}^{2}+(1-x-y)\Lambda _{1}^{2} 
\right \}^{2}} 
\nonumber \\ 
& & \hskip1cm +\frac{(4\xi _{0} -2)y^{2}m_{p}^{2}}{\left \{ 
y^{2}m_{p}^{2}+x\Lambda _{0}^{2}+(1-x-y)\Lambda _{1}^{2} 
\right \}^{3}} 
\Bigg ],
\end{eqnarray} 
and  $\xi _{0}$ is either 1/2 or 2. 
A similar expression is used to evaluate (\ref{eq:dspdtp}).

The evaluation of (\ref{eq:e(p)}) goes similarly, 
starting from \begin{eqnarray} 
\frac{i}{16\pi ^{2}}{\cal E}^{(p)}&=&\frac{1}{m_{p}^{2}} 
\int \frac{d^{4}k}{(2\pi )^{4}}\frac{(k\cdot p_{2})}{k^{2}} 
\frac{1}{(p_{2}-k)^{2}-m_{p}^{2}}  
\nonumber \\
& & \times 
F_{W}(k^{2}) 
\left \{ F_{1}^{(p)}(k^{2})+F_{2}^{(p)}(k^{2}) 
\right \}. 
\end{eqnarray} 
Our numerical computation employs the Feynman integral 
\begin{eqnarray} 
\int \frac{d^{4}k}{(2\pi )^{4}}\frac{(k\cdot p_{2})}{k^{2}} 
\frac{1}{(p_{2}-k)^{2}-m_{p}^{2}} 
\frac{1}{(1-k^{2}/\Lambda _{0}^{2})^{2}(1-k^{2}/\Lambda _{1}^{2})^{2}} 
\\ 
=\frac{\Lambda _{0}^{4}\Lambda _{1}^{4}}{2}\left \{ 
h(1, \Lambda _{0}, \Lambda _{1})-h(0, \Lambda _{0}, \Lambda _{1}) 
\right \}, 
\end{eqnarray} 
where 
\begin{eqnarray} 
h(\xi , \Lambda _{0}, \Lambda _{1})&=&\frac{i}{16\pi ^{2}} 
\int _{0}^{1} dx \int _{0}^{1-x} 
dy x(1-x-y)
\nonumber \\
& & \times \frac{-2}{ 
\left \{ y^{2}\xi ^{2}m_{p}^{2}+x\Lambda _{0}^{2}+(1-x-y)\Lambda _{1}^{2} 
\right \}^{3}}. 
\end{eqnarray} 
Likewise, ${\cal E}^{(n)}$ is computed 
via the same type of parameter integrals.

\end{document}

%% file: qed.tex
\unitlength 0.1in 
\begin{picture}(40.55,38.77)(1.00,-39.60) 
\special{pa 279 449}%
\special{pa 557 1006}%
\special{fp}%
\special{pa 279 1716}%
\special{pa 584 1239}%
\special{fp}%
\put(11.1200,-4.4100){\makebox(0,0)[lb]{$e^{+}$}}%
\put(1.5300,-18.3200){\makebox(0,0)[lb]{$p$}}%
\put(11.4700,-19.0400){\makebox(0,0)[lb]{$\bar \nu _{e}$}}%
\put(2.2600,-4.0500){\makebox(0,0)[lb]{$n$}}%
\special{sh 0.300}%
\special{ar 700 1105 177 178  0.0000000 6.2831853}%
\special{pa 1067 2451}%
\special{pa 1344 3008}%
\special{fp}%
\special{pa 1067 3717}%
\special{pa 1371 3241}%
\special{fp}%
\put(18.4600,-24.7800){\makebox(0,0)[lb]{$e^{+}$}}%
\put(9.8600,-38.4300){\makebox(0,0)[lb]{$p$}}%
\put(19.2700,-39.0600){\makebox(0,0)[lb]{$\bar \nu _{e}$}}%
\put(9.5000,-24.2500){\makebox(0,0)[lb]{$n$}}%
\special{sh 0.300}%
\special{ar 1488 3106 177 178  0.0000000 6.2831853}%
\special{pa 2616 2461}%
\special{pa 2893 3017}%
\special{fp}%
\special{pa 2616 3726}%
\special{pa 2919 3251}%
\special{fp}%
\put(33.9400,-24.4300){\makebox(0,0)[lb]{$e^{+}$}}%
\put(25.3500,-38.7000){\makebox(0,0)[lb]{$p$}}%
\put(34.7500,-38.8800){\makebox(0,0)[lb]{$\bar \nu _{e}$}}%
\put(25.1700,-24.2500){\makebox(0,0)[lb]{$n$}}%
\special{sh 0.300}%
\special{ar 3037 3116 177 177  0.0000000 6.2831853}%
\special{pa 1899 431}%
\special{pa 2177 988}%
\special{fp}%
\special{pa 1899 1697}%
\special{pa 2204 1221}%
\special{fp}%
\put(27.5900,-4.4100){\makebox(0,0)[lb]{$e^{+}$}}%
\put(18.2800,-18.3200){\makebox(0,0)[lb]{$p$}}%
\put(27.5900,-18.8600){\makebox(0,0)[lb]{$\bar \nu _{e}$}}%
\put(18.6400,-3.9600){\makebox(0,0)[lb]{$n$}}%
\special{sh 0.300}%
\special{ar 2320 1087 178 178  0.0000000 6.2831853}%
\special{pa 3269 449}%
\special{pa 3547 1006}%
\special{fp}%
\special{pa 3269 1716}%
\special{pa 3573 1239}%
\special{fp}%
\put(41.2000,-4.4900){\makebox(0,0)[lb]{$e^{+}$}}%
\put(32.0600,-18.6800){\makebox(0,0)[lb]{$p$}}%
\put(41.5500,-19.0400){\makebox(0,0)[lb]{$\bar \nu _{e}$}}%
\put(32.0600,-4.2300){\makebox(0,0)[lb]{$n$}}%
\special{sh 0.300}%
\special{ar 3690 1105 178 178  0.0000000 6.2831853}%
\put(6.1900,-20.9200){\makebox(0,0)[lb]{(v)}}%
\put(21.7700,-41.3000){\makebox(0,0)[lb]{(b)}}%
\put(29.6500,-21.0100){\makebox(0,0)[lb]{(s)}}%
\special{ar 664 1087 421 422  0.8235691 0.9659179}%
\special{ar 664 1087 421 422  1.0513272 1.1936759}%
\special{ar 664 1087 421 422  1.2790852 1.4214339}%
\special{ar 664 1087 421 422  1.5068432 1.6491919}%
\special{ar 664 1087 421 422  1.7346012 1.8769499}%
\special{ar 664 1087 421 422  1.9623592 2.1047079}%
\special{ar 664 1087 394 440  5.7091191 5.8530040}%
\special{ar 664 1087 394 440  5.9393349 6.0832198}%
\special{ar 664 1087 394 440  6.1695507 6.3134356}%
\special{ar 664 1087 394 440  6.3997666 6.5436515}%
\special{ar 664 1087 394 440  6.6299824 6.7658724}%
\special{ar 3958 791 159 158  5.4604124 5.8389613}%
\special{ar 3958 791 159 158  6.0660907 6.4446395}%
\special{ar 3958 791 159 158  6.6717689 7.0503178}%
\special{ar 3958 791 159 158  7.2774471 7.6559960}%
\special{ar 2096 1500 165 165  2.5467051 2.9103415}%
\special{ar 2096 1500 165 165  3.1285233 3.4921597}%
\special{ar 2096 1500 165 165  3.7103415 4.0739778}%
\special{ar 2096 1500 165 165  4.2921597 4.6557960}%
\special{ar 2096 1500 165 165  4.8739778 4.9264933}%
\special{pa 1523 2415}%
\special{pa 1747 2918}%
\special{da 0.070}%
\special{pa 2562 2963}%
\special{pa 2785 3466}%
\special{da 0.070}%
\put(5.5700,-16.4300){\makebox(0,0)[lb]{$\gamma $}}%
\put(14.7000,-27.3000){\makebox(0,0)[lb]{$\gamma $}}%
\put(18.2800,-14.0100){\makebox(0,0)[lb]{$\gamma $}}%
\put(25.0800,-33.3200){\makebox(0,0)[lb]{$\gamma $}}%
\put(41.2800,-9.0800){\makebox(0,0)[lb]{$\gamma $}}%
\put(1.0000,-15.9000){\makebox(0,0)[lb]{$(p_{2})$}}%
\put(1.0000,-2.7000){\makebox(0,0)[lb]{$(p_{1})$}}%
\put(10.8500,-2.5300){\makebox(0,0)[lb]{$(\ell )$}}%
\special{pa 1130 467}%
\special{pa 879 1096}%
\special{pa 1147 1734}%
\special{pa 1147 1734}%
\special{pa 1147 1734}%
\special{fp}%
\special{pa 4120 512}%
\special{pa 3868 1140}%
\special{pa 4138 1778}%
\special{pa 4138 1778}%
\special{pa 4138 1778}%
\special{fp}%
\special{pa 2750 467}%
\special{pa 2499 1096}%
\special{pa 2768 1734}%
\special{pa 2768 1734}%
\special{pa 2768 1734}%
\special{fp}%
\special{pa 3466 2469}%
\special{pa 3215 3098}%
\special{pa 3484 3735}%
\special{pa 3484 3735}%
\special{pa 3484 3735}%
\special{fp}%
\special{pa 1917 2496}%
\special{pa 1667 3124}%
\special{pa 1935 3762}%
\special{pa 1935 3762}%
\special{pa 1935 3762}%
\special{fp}%
\special{pa 2195 3439}%
\special{pa 2195 3439}%
\special{fp}%
\end{picture}%

%% file: wz.tex
\unitlength 0.1in
\begin{picture}(47.40,14.44)(0.40,-15.36)
%
\special{pn 13}%
\special{pa 116 1528}%
\special{pa 378 916}%
\special{pa 116 304}%
\special{pa 116 304}%
\special{pa 116 304}%
\special{fp}%
%
\special{pn 13}%
\special{pa 1012 313}%
\special{pa 741 924}%
\special{pa 995 1536}%
\special{pa 995 1536}%
\special{pa 995 1536}%
\special{fp}%
\put(0.4800,-2.7100){\makebox(0,0)[lb]{$d$}}%
\put(0.4000,-16.6200){\makebox(0,0)[lb]{$u$}}%
\put(8.6800,-2.7900){\makebox(0,0)[lb]{$e^{+}$}}%
\put(8.5100,-16.7000){\makebox(0,0)[lb]{$\bar \nu _{e}$}}%
%
\special{pn 13}%
\special{pa 370 916}%
\special{pa 733 924}%
\special{da 0.070}%
%
\special{pn 13}%
\special{pa 1350 1519}%
\special{pa 1612 908}%
\special{pa 1350 296}%
\special{pa 1350 296}%
\special{pa 1350 296}%
\special{fp}%
%
\special{pn 13}%
\special{pa 2245 304}%
\special{pa 1975 916}%
\special{pa 2228 1528}%
\special{pa 2228 1528}%
\special{pa 2228 1528}%
\special{fp}%
\put(12.8200,-2.6200){\makebox(0,0)[lb]{$d$}}%
\put(12.7400,-16.5300){\makebox(0,0)[lb]{$u$}}%
\put(21.0200,-2.7100){\makebox(0,0)[lb]{$e^{+}$}}%
\put(20.8500,-16.6200){\makebox(0,0)[lb]{$\bar \nu _{e}$}}%
%
\special{pn 13}%
\special{pa 1603 908}%
\special{pa 1966 916}%
\special{da 0.070}%
%
\special{pn 13}%
\special{pa 2617 1519}%
\special{pa 2879 908}%
\special{pa 2617 296}%
\special{pa 2617 296}%
\special{pa 2617 296}%
\special{fp}%
%
\special{pn 13}%
\special{pa 3513 304}%
\special{pa 3242 916}%
\special{pa 3496 1528}%
\special{pa 3496 1528}%
\special{pa 3496 1528}%
\special{fp}%
\put(25.4900,-2.6200){\makebox(0,0)[lb]{$d$}}%
\put(25.4100,-16.5300){\makebox(0,0)[lb]{$u$}}%
\put(33.6900,-2.7100){\makebox(0,0)[lb]{$e^{+}$}}%
\put(33.5200,-16.6200){\makebox(0,0)[lb]{$\bar \nu _{e}$}}%
%
\special{pn 13}%
\special{pa 2870 908}%
\special{pa 3234 916}%
\special{da 0.070}%
%
\special{pn 13}%
\special{pa 3884 1519}%
\special{pa 4146 908}%
\special{pa 3884 296}%
\special{pa 3884 296}%
\special{pa 3884 296}%
\special{fp}%
%
\special{pn 13}%
\special{pa 4780 304}%
\special{pa 4510 916}%
\special{pa 4763 1528}%
\special{pa 4763 1528}%
\special{pa 4763 1528}%
\special{fp}%
\put(38.1700,-2.6200){\makebox(0,0)[lb]{$d$}}%
\put(38.0800,-16.5300){\makebox(0,0)[lb]{$u$}}%
\put(46.3600,-2.7100){\makebox(0,0)[lb]{$e^{+}$}}%
\put(46.1900,-16.6200){\makebox(0,0)[lb]{$\bar \nu _{e}$}}%
%
\special{pn 13}%
\special{pa 4138 908}%
\special{pa 4501 916}%
\special{da 0.070}%
\put(4.2000,-10.8400){\makebox(0,0)[lb]{$W$}}%
\put(16.7100,-8.7400){\makebox(0,0)[lb]{$W$}}%
\put(28.6200,-10.9200){\makebox(0,0)[lb]{$W$}}%
\put(43.3200,-10.6700){\makebox(0,0)[lb]{$W$}}%
%
\special{pn 13}%
\special{ar 538 1125 592 587  4.2048534 4.3066345}%
\special{ar 538 1125 592 587  4.3677032 4.4694844}%
\special{ar 538 1125 592 587  4.5305531 4.6323343}%
\special{ar 538 1125 592 587  4.6934030 4.7951842}%
\special{ar 538 1125 592 587  4.8562529 4.9580340}%
\special{ar 538 1125 592 587  5.0191027 5.1208839}%
\special{ar 538 1125 592 587  5.1819526 5.2733859}%
%
\special{pn 13}%
\special{ar 1781 740 506 502  0.9710857 1.0901333}%
\special{ar 1781 740 506 502  1.1615619 1.2806095}%
\special{ar 1781 740 506 502  1.3520381 1.4710857}%
\special{ar 1781 740 506 502  1.5425142 1.6615619}%
\special{ar 1781 740 506 502  1.7329904 1.8520381}%
\special{ar 1781 740 506 502  1.9234666 2.0425142}%
\special{ar 1781 740 506 502  2.1139428 2.1515951}%
%
\special{pn 13}%
\special{ar 2541 1025 541 520  5.0484698 5.1615706}%
\special{ar 2541 1025 541 520  5.2294311 5.3425320}%
\special{ar 2541 1025 541 520  5.4103925 5.5234933}%
\special{ar 2541 1025 541 520  5.5913538 5.7044547}%
\special{ar 2541 1025 541 520  5.7723152 5.8854160}%
\special{ar 2541 1025 541 520  5.9532766 5.9653410}%
%
\special{pn 13}%
\special{ar 3529 882 473 419  1.8479336 1.9824628}%
\special{ar 3529 882 473 419  2.0631803 2.1977094}%
\special{ar 3529 882 473 419  2.2784269 2.4129561}%
\special{ar 3529 882 473 419  2.4936736 2.6282027}%
\special{ar 3529 882 473 419  2.7089202 2.8434493}%
\special{ar 3529 882 473 419  2.9241668 2.9608333}%
%
\special{pn 13}%
\special{ar 3960 841 364 410  0.2750900 0.4301288}%
\special{ar 3960 841 364 410  0.5231521 0.6781908}%
\special{ar 3960 841 364 410  0.7712141 0.9262528}%
\special{ar 3960 841 364 410  1.0192761 1.1743148}%
\special{ar 3960 841 364 410  1.2673381 1.4223769}%
%
\special{pn 13}%
\special{ar 4679 1042 372 486  3.4875599 3.6274201}%
\special{ar 4679 1042 372 486  3.7113361 3.8511963}%
\special{ar 4679 1042 372 486  3.9351124 4.0749725}%
\special{ar 4679 1042 372 486  4.1588886 4.2987487}%
\special{ar 4679 1042 372 486  4.3826648 4.5225250}%
\special{ar 4679 1042 372 486  4.6064410 4.6823096}%
\put(5.0500,-4.8000){\makebox(0,0)[lb]{$\gamma $}}%
\put(40.3600,-13.7700){\makebox(0,0)[lb]{$\gamma $}}%
\put(5.1300,-6.8100){\makebox(0,0)[lb]{$Z$}}%
\put(16.9600,-13.9400){\makebox(0,0)[lb]{$Z$}}%
\put(30.2300,-7.5700){\makebox(0,0)[lb]{$Z$}}%
\put(41.6300,-13.5200){\makebox(0,0)[lb]{$Z$}}%
\end{picture}%